\documentclass{aa}  
\usepackage{natbib}
\usepackage{graphicx}

\usepackage{txfonts}
\begin{document}
   \title{Probing high-redshift quasars with ALMA}

   \subtitle{I. Expected observables and potential number of sources}

   \author{Dominik R. G. Schleicher
          \inst{1,2}
          \and
          Marco Spaans \inst{3}\fnmsep%\thanks{Just to show the usage
   %       of the elements in the author field
          \and
          Ralf S. Klessen \inst{4}
          }

  % \offprints{G. Wuchterl}

   \institute{Leiden Observatory, P.O. Box 9513, NL-2300 RA Leiden, the Netherlands\\
              \email{schleicher@strw.leidenuniv.nl}
              \and
              ESO, Karl-Schwarzschild-Strasse 2, 85748 Garching bei M\"unchen, Germany
         \and
             Kapteyn Astronomical Institute, University of Groningen, P.O. Box 800, 9700 AV, Groningen, the Netherlands\\
             \email{spaans@astro.rug.nl}
             \and
             Zentrum f\"ur Astronomie der Universit\"at Heidelberg, Institut f\"ur Theoretische Astrophysik, Albert-Ueberle-Str. 2, 69120 Heidelberg, Germany\\
             \email{rklessen@ita.uni-heidelberg.de}
             }

   \date{Received September 15, 1996; accepted March 16, 1997}

% \abstract{}{}{}{}{} 
% 5 {} token are mandatory
 
  \abstract
%   {Observations in the mm and sub-mm regime are valuable probes for the physical conditions of starburst and quasar host galaxies. With the upcoming Atacama Large Millimeter Array (ALMA), it will be possible to probe high-redshift galaxies with unprecedented resolution and sensitivity, and at unprecedented redshifts. Due to upcoming surveys which will be performed for instance by JWST, observational targets may become available even at $z=10$. }
{}
   {We explore how ALMA observations can probe high-redshift galaxies in unprecedented detail. We discuss the main observables that are excited by the large-scale starburst, and formulate expectations for the chemistry and the fluxes in the center of active galaxies, in which chemistry may be driven by the absorption of X-rays. We estimate the expected number of sources at high redshift to infer whether an ALMA deep field may find a reasonable number. {As a specific example for the complex interpretation of sub-mm line observations, we analyze the recently detected $z=6.42$ quasar, for which a number of different line fluxes is already available. We note that our diagnostics may also be valuable for future observations in the local universe with space-borne instruments like on SPICA or FIRI.}}
   {To estimate the observables from the starburst, we check which emission from the starburst ring of the nearby Seyfert 2 galaxy NGC~1068 falls into the ALMA bands if the galaxy were placed at $z=8$.  We estimate the sizes of the central X-ray dominated region based on a semi-analytic model, and employ a detailed 1D approach for the chemistry in X-ray irradiated molecular clouds to evaluate the chemistry and the expected line emission under these conditions. We make use of pre-existing chemistry calculations in X-ray dominated regions to show the dependence of different line fluxes on X-ray luminosity, cloud density and cloud column density. We use theoretical models for the high-$z$ black hole population and the local SMBH density to estimate the number of sources at higher redshift.}
   {We show that a number of different fine-structure lines may be used to probe the starburst component of high-redshift quasars in considerable detail, providing specific information on the structure of these galaxies by several independent means. We show that the size of the central X-ray dominated region is of the order of a few hundred parsec, and we provide detailed predictions for the expected fluxes in CO, [CII] and [OI]. While the latter fine-structure lines quickly become optically thick and depend mostly on the strength of the X-ray source, the rotational CO lines have a non-trivial dependence on these parameters. We compare our models to XDRs observed in NGC~1068 and APM~08279 and find that the observed emission can indeed be explained with these models. Depending on the amount of X-ray flux, the CO line intensities may rise continuously up to the (17-16) transition. {A measurement of such high-$J$ lines allows one to distinguish observationally between XDRs and PDRs. For the recently observed $z=6.42$ quasar, we show that the collected fluxes cannot be interpreted in terms of a single gas component. We find indications for the presence of a dense warm component in active star forming regions and a low-density component in more quiescent areas.} Near $z=6$, an ALMA deep field may find roughly one source per arcmin$^2$. At higher redshift, one likely has to rely on other surveys like JWST to find appropriate sources.}
%   {We show that ALMA will be valuable to improve our understanding of the PDR component in high-redshfit galaxies, and may for the first time attempt to detect emission from the central X-ray dominated regions in these sources. In these regions, we expect a significantly different type of chemistry than during the starburst, reflected in high temperatures and still reasonable CO abundances. A detection of the high-$J$ CO lines would therefore provide direct evidence for the presence of such reasons. Strong emission is also expected in fine-structure lines like [CII] and [OI]. Observations in NGC~1068 and APM~08279 are consistent with these models. An ALMA deep field may be possible to achieve near $z=6$, while it is necessary to rely on other surveys at higher redshift.}
{}

   \maketitle
%
%________________________________________________________________

\section{Introduction}
The recent detection of kpc-scale star-forming structures at $z=6.42$ through the detection of [CII] emission \citep{Walter09} and emission in various rotational CO lines \citep{Riechers09} confirms the importance of mm/sub-mm observations to infer gas distribution and dynamics in quasar host galaxies. Emission in CO and the continuum in high-redshift quasars have also been reported by \citet{Omont96}, \citet{Carilli02}, \citet{Walter04}, \citet{Klamer05}, \citet{Weiss05}, \citet{Maiolino07}, \citet{Walter07}, \citet{Weiss07}, \citet{Riechers08a} and \citet{Riechers08b}. 

With the upcoming mm/sub-mm telescope ALMA\footnote{http://www.eso.org/sci/facilities/alma/}, it will be possible to probe such structures in even more detail, due to its significantly improved sensitivity, angular and spectral resolution. From a theoretical point of view, it is therefore interesting to speculate what ALMA might observe in the center of such host galaxies. {Within the next ten years, we further expect  the advent of SPICA\footnote{http://www.ir.isas.jaxa.jp/SPICA/index.html} and FIRI\footnote{http://sci.esa.int/science-e/www/object/index.cfm?fobjectid=40090}, which will probe the universe in the mid- or far-IR regime, respectively. These telescopes will be able to apply in the local universe what we suggest as high-redshift diagnostics for ALMA.}

{Before speculating what ALMA may see in the centers of high-redshift galaxies, we turn our attention to} the properties of molecular clouds in the central molecular zone (CMZ) of the Milky Way. Studies employing H$_3^+$ and CO lines indicate the presence of high-temperature ($T\sim250$~K) and low-density ($n\sim100$~cm$^{-3}$) gas \citep{Oka05}. NH$_3$ observations by \citet{Nagayama07} confirm the presence of warm molecular clouds, with temperatures mostly between $20-80$~K. The presence of a $120$~pc star-forming ring was infered by CO observations, indicating typical densities of $10^{3.5-4}$~cm$^{-3}$ and kinetic temperatures of $20-35$~K \citep{Nagai07}. These temperatures were derived under the conservative assumption of a beam filling factor $1$, while smaller filling factors would give rise to higher temperatures.  

In the centers of active galaxies,  the supermassive black hole will emit radiation in a broad range of frequencies. Particularly interesting is the emission of X-rays, as those photons can penetrate molecular clouds even at high column densities. The resulting cloud temperatures range from a hundred K up to $1000$~K \citep{Lepp96, Maloney96, Meijerink05, Meijerink07}. Such X-ray dominated regions have been reported for instance in NGC~1068 by \citet{Galliano03}. Due to the high ambient pressure, higher-density clouds may form due to the thermal instability \citep{Wada07}.

At high redshift, resolution of present-day telescopes is generally not sufficient to resolve the central regions of quasar host galaxies. Under exceptional circumstances, this is however possible. Indeed, highly excited high-$J$ CO and HCN line emission was found in APM~08279+5255 by \citet{Weiss07}. As this galaxy is  gravitationally lensed, it was possible to measure fluxes from the central X-ray dominated region that are usually beam-diluted. While the CO line fluxes usually rise up to the CO~(5-4) transition and then decrease, they continue to rise in this system up to the CO~(10-9) transition, providing clear evidence for the presence of warm gas in the central region. {Similar results have also been obtained for the Cloverleaf quasar \citep{Bradford09}.}

Motivated by these results, we study in more detail the possibility to probe high-redshift quasar host galaxies with ALMA, and in particular their central regions. In \S~\ref{heat}, we review the main heating mechanisms that may be present in high-redshift quasars, and discuss their influence on the chemistry. In \S~\ref{PDR}, we give a brief summary on the main PDR observables, which are already used to probe galaxies at high redshift. As a specific example, we calculate the expected fluxes for the Seyfert 2 galaxy NGC~1068 if it were located at $z=8$. Our expectations for the central X-ray dominated regions (XDRs) are formulated in \S~\ref{XDR} based on detailed chemical models including $\sim50$ species and several thousand reactions. Evidence for X-ray dominated regions at different redshifts is reviewed and discussed in \S~\ref{evidence}. On this basis, we assess the prospects for finding new sources in an ALMA deep field in \S~\ref{deep_field}.  We conclude in \S~\ref{conclusions}. In summary, this paper provides a basic set of predictions concerning observations of high-redshift quasars with ALMA. In a companion paper, we plan to provide diagnostics based on the observed line fluxes that will  allow one to infer physical properties such as the star formation rate or the X-ray luminosity based on the observed line emission.

\section{Chemistry in high-redshift quasars}\label{heat}

\begin{figure}[t]
\includegraphics[scale=0.45]{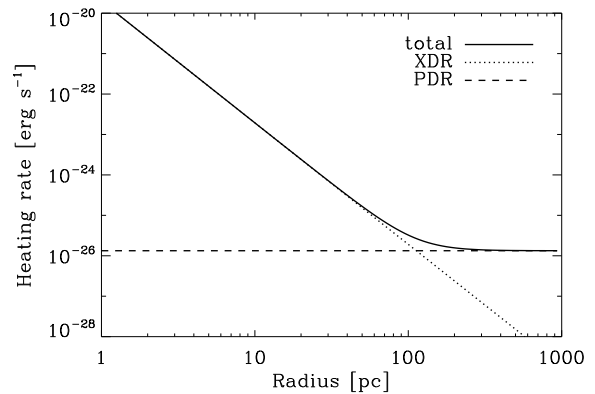}\caption{The heating rates per hydrogen atom due to X-ray absorption (XDR contribution), absorption of soft UV-photons (PDR contribution) and total heating rate as a function of radius. The calculation assumes a $10^7\ M_\odot$ black hole, with $3\%$ of its Eddington luminosity being emitted in a hard spectral component between $1$ and $100$~keV with a spectral slope of $-1$. The strength of the soft UV radiation field is taken as $G_0=10$ in Habing units and the effective density $n_{\mathrm{eff}}=10^5$~cm$^{-3}$. For such a configuration, the XDR contribution clearly dominates within the central $100$~pc.}
\label{fig:heating}
\end{figure}

\begin{figure}[t]
\includegraphics[scale=0.45]{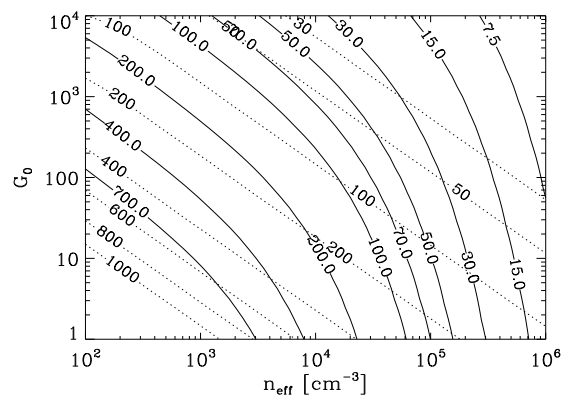}
\caption{The expected size of the X-ray dominated region in pc, for a black hole with $10^7\ M_\odot$ with a spectral slope of $-1$, as a function of the soft UV radiation field $G_0$ in Habing units and the effective number density $n_{\mathrm{eff}}$. The straight lines assume the power-law between $1$ and $5$~keV, while the dotted lines assume it between $1$ and $100$~keV.}
\label{fig:models}
\end{figure}
Emission from molecular clouds in active galaxies can be excited by a variety of different mechanisms. Mechanical feedback may be important locally, in particular in the presence of shocks or outflows \citep{Loenen08, Papadopoulos08}. In addition, there is radiation in a broad range of frequencies, both from the starburst and the supermassive black hole. UV emission generally gives rise to compact HII regions at temperatures of $\sim10^4$~K in which molecules are completely dissociated and emission is mostly by Lyman $\alpha$ and various fine-structure lines. Such photons are however absorbed by relatively small column densities, and stellar HII regions never become larger than a few pc, considerably smaller than the scales of interest here. 

Soft UV-photons have smaller cross sections and may penetrate larger columns. In the presence of a starburst, they are therefore the dominant driver of the molecular cloud chemistry \citep{Hollenbach99}. While their heating efficiency is low $(0.1-0.3\%)$, they are very efficient in dissociating molecules. In such photon-dominated regions (PDRs), one generally expects somewhat enhanced cloud temperatures, with most of the emission in fine-structure lines, as CO is efficiently dissociated. A detailed review of these processes is given by \citet{Meijerink05}.

X-ray photons have even smaller cross sections than the soft-UV photons, and {can thus penetrate larger columns. Specifically, a 1~keV photon penetrates a typical column of $2\times10^{22}$~cm$^{-2}$, a 10~keV photon penetrates  $4\times10^{25}$~cm$^{-2}$ and a 100~keV photon  $9\times10^{30}$~cm$^{-2}$. For this reason, X-rays can keep molecular clouds at high temperatures even at high column densities}. They have high heating efficiencies of the order of $30\%$, and are inefficient in the dissociation of molecules. A fraction of them may however be reprocessed by the gas and converted in soft-UV photons, which may lead to some molecular dissociation. Detailed reviews are given by \citet{Maloney96, Lepp96, Meijerink05}. Therefore, X-ray absorption drives a completely different type of chemistry, and may potentially result in temperatures up to $1000$~K. The fraction of molecules, in particular CO, can be very high.

To estimate the potential extent of such a central X-ray dominated region, we employ a toy model that takes into account the heat input from the starburst and from the X-ray emission of the supermassive black hole. We assume here an {axisymmetric situation with a central supermassive black hole and an extended molecular disk in the host galaxy.} The radiation from the SMBH will consist of a soft and a hard component. The soft component is easily absorbed at the edge of the molecular clouds and can be neglected for column densities of $10^{22}$~cm$^{-2}$ and above. The X-ray photons, on the other hand, penetrate deeply into the molecular disk and excite emission there. For the X-ray photons, we adopt a power-law spectrum for frequencies larger than $1$~keV and use the cross sections given by \citet{Verner95}. For soft photons from the starburst, we adopt a typical frequency of $10$~eV and a cross section of $2.78\times10^{-22}$~cm$^{-2}$ \citep{Meijerink05}. Typical heating efficiencies are $30\%$ in XDRs and $0.3\%$ in PDRs. We assume that the radiation field from the starburst is roughly constant within the central region. The X-ray radiation field, on the other hand, will be geometrically diluted and partially shielded by the gas. To calculate the attenuation of X-rays, we introduce an effective density for the central region, which is given as
\begin{equation}
n_{\mathrm{eff}}=\alpha 10^5\ \mathrm{cm}^{-3}+(1-\alpha) 10^2\ \mathrm{cm}^{-3},
\end{equation}
where $\alpha$ is the volume-filling factor of dense clouds, $10^5$~cm$^{-3}$ is a typical cloud density and $10^2$~cm$^{-3}$ a typical density of the atomic medium. We adopt now a specific reference case for which we evaluate the heating rates and the expected size of the XDR. The X-ray emission depends essentially on the product of black hole mass $M_{\mathrm{BH}}$, the Eddington ratio $\lambda_E$, and the fraction $f_X$ of the total luminosity going into the hard spectrum. We assume a modest black hole mass $M_{BH}=10^7\ M_\odot$, an Eddington ratio $\lambda_E=30\%$, which lies in the typical range of measured Eddington ratios for high-redshift AGN \citep{Shankar04, Kollmeier06, Shankar09}, and a fraction of $f_X=10\%$ of the total luminosity emitted in X-rays. For the X-ray spectrum, we adopt a frequency range between $1$ and $100$~keV with a spectral slope of $-1$.

We further assume that the starburst produces a soft UV-radiation field of $G_0=10$ (Habing units) and an effective density of $n_{\mathrm{eff}}=10^5$~cm$^{-3}$. Such an effective density corresponds to a volume-filling factor of order $1$ and is therefore an upper limit. In real AGN, the effective density may be smaller in the central region, implying less attenuation of the X-rays and therefore a larger X-ray dominated region. On the other hand, the parameter $G_0$ may vary as well and depend on the strength of the ongoing starburst. For this scenario, the expected heating rates per hydrogen atom are given in Fig.~\ref{fig:heating}. 

To explore the parameter dependence in more detail, we now consider the effective density and the strength of the starburst radiation field as free parameters and check how they influence the size of the XDR, keeping black hole mass, Eddington ratio and the luminosity fraction in the hard component as specified above. We consider two cases, one with the hard component between $1$ and $5$~keV (case A), and one with the hard component between $1$ and $100$~keV (case B). The results for the XDR size are given in Fig.~\ref{fig:models}. In case A shielding effects can be clearly recognized and the size of the XDR depends more on the effective density than on the strength of the starburst.  {As typical molecular cloud densities in these environments are $\sim10^5$~cm$^{-3}$ and the filling factor is of order $1\%$, we can expect an average density of $10^3$~cm$^{-3}$. At its largest baseline, ALMA can even resolve spatial scales of $\sim30$~pc at $z=5$, and should thus resolve the corresponding XDRs.}

%much stronger than $G_0=100$. One should note that for the case of relatively small effective densities of $\sim10^3$~cm$^{-3}$, shielding is not effective, and the XDR may reach out further. In this low-density case, one obtains larger XDRs in case A with a smaller spectral range.

Simulations by \citet{Wada09} for the clumpy medium at scales of $\sim30$~pc indicate a volume-filling factor $\alpha\sim0.03$. Models by \citet{Galliano03} for NGC~1068, on scales of a few hundred parsec, indicate a volume-filling factor of $0.01$, which still leads to a surface-filling factor of order 1. For the central $200$~pc of our galaxy, values of $\alpha\sim0.1-0.01$ have been suggested \citep{McCall99, Oka05}. For a large volume-filling factor, we expect a somewhat smaller XDR due to the attenuation of X-rays. At the same time, however, this region will consist of a large number of clouds that are highly excited. For smaller clumping factors, the number densitiy of clouds will be reduced, but the size of the XDR increased.

Of course, the approach used here is only approximate, as the attenuation may depend on the direction and be stronger in some directions and weaker in others. However, these order-of-magnitude estimates should still be applicable for a broad range of conditions and also hold in cases of spherical rather than flattened structures. An implicit assumption of the model is that we average over sufficiently large scales where the effective density provides a good approximation with respect to X-ray attenuation. For more clumpy structures, the XDR would be more inhomogeneous and reach out further along the low-density regions. {In case the large-scale clumpiness is considerable, so that molecular clouds do no longer fill the projected surface area in the beam, our model predictions need to be corrected with the corresponding area-filling factor.}

\section{Observables in the PDR}\label{PDR}

In photon-dominated regions, soft UV-photons from the starburst provide some heat input for molecular clouds, in particular at low column densities, and are efficient in dissociating molecules like H$_2$ or CO. The main coolants in this regime are therefore fine-structure lines of [CII] and [OI]. Depending on the strength of the radiation field, some flux may also be emitted in molecules like CO or HCN, in particular in the low-lying rotational transitions, and there are further fine-structure lines that may contribute as well. As mentioned in the introduction, there are plenty of observations that studied emission from starburst galaxies with present-day sub-mm telescopes \citep[e.g.][]{Omont96, Carilli02, Walter04, Klamer05, Weiss05, Maiolino07, Walter07,  Riechers08a, Riechers08b, Greve09,Riechers09,Walter09}.

\begin{table}[htdp]
\begin{center}
\begin{tabular}{llll}
Observable & $\lambda$~[$\mathrm{\mu}$m] & $\varphi~[mJy]$ & min. Redshift \\
\hline
%high-$J$ \COI lines  &$\sim300$ &  $0.04-1$   &  $z>5$ \\
%high-$J$ HCN lines  & $\sim300$ &  $0.004-0.1$   & $z>4$ \\
%high-$J$ HCO$^+$   & $\sim300$  &  $0.004-0.1$   & $z>4$ \\
$[$O I$]$~$\ ^3P_1\rightarrow^3P_2$& $63.2$ & $\sim0.18$   &  $z>5.7$ \\
$[$O~III$]$~$^3P_1\rightarrow^3P_0$& $51.8$ & $\sim0.18$  &  $z>3.8$ \\
$[$N~II$]$~$^3P_2\rightarrow^3P_1$ & $121.9$ &  $\sim0.07$  &  $z>2.5$ \\
$[$O I$]$~$\ ^3P_0\rightarrow^3P_1$ & $145.5$  & $\sim0.03$   &  $z>1.9$ \\
$[$C II$]$~$\ ^2P_{3/2}\rightarrow^2P_{1/2}$  & $157.7$ & $\sim0.7$   &  $z>1.7$ \\
$[$S III$]$~$\ ^3P_1\rightarrow^3P_0$ & $33.5$ & $\sim1.5$   &  $z>11.1$ \\
$[$Si II$]$~$\ ^2P_{3/2}\rightarrow^2P_{1/2}$ & $34.8$ & $\sim0.06$   &  $z>11.1$ \\
%dust continuum  & $\sim500$ & $\sim0.05$ &  $z>0$  \\
\hline
\end{tabular}
\end{center}
\caption{The main observables for ALMA in PDRs at high redshift. This specific example assumes a galaxy with a starburst as in NGC~1068 placed at $z=8$. We also give the minimal redshift from which the lines would be redshifted into the ALMA bands.}
\label{tab:observables}
\end{table}%

With the upcoming sensitivity of ALMA, we expect that such PDRs can be probed in more detail as well. Therefore, bright lines like the [CII]~$158$~$\mu$m line can be detected at higher significance, allowing higher spectral resolution and probing the velocity structure of the gas in more detail. Also, weaker lines may be detected as well, probing gas at different densities and providing additional information on the chemical conditions. 

\begin{table}[htdp]
\begin{center}
\begin{tabular}{cccccc}
band & freq. [GHz]  & $\theta{\mathrm{res}}$~[''] & $S_c$~[mJy] & $S_l$~[mJy] & $\theta_{\mathrm{beam}}$~[''] \\
\hline
3& $84-116$  & $0.034$ & $0.019$ & $0.163$ & $56$ \\
 4&$125-169$ & $0.023$ & $0.023$ & $0.174$& $48$ \\
 5&$163-211$ & $0.018$ & $0.298$ & $2.63$& $35$ \\
 6&$211-275$ & $0.014$ & $0.039$ & $0.225$& $27$ \\
 7&$275-373$ & $0.011$ & $0.077$ & $0.372$& $18$ \\
 8&$385-500$ & $0.008$ & $0.143$ & $0.620$& $12$ \\
 9&$602-720$ & $0.005$ & $0.232$ & $0.813$& $9$ \\
\hline
\end{tabular}
\end{center}
\caption{Frequency range, angular resolution $\theta_{\mathrm{res}}$ at the largest baseline, line sensitivity $S_l$ {for a linewidth of $300$~km/s} and continuum sensitivity $S_c$  {for $3\sigma$ detection in one hour} of integration time  and primary beam size $\theta_{\mathrm{beam}}$. 3 more bands might be added in the future, band $1$ around $40$~GHz, band $2$ around $80$~GHz and band $10$ around $920$~GHz, which will have similar properties as the neighbouring bands.}
\label{tab:ALMA}
\end{table}%

To obtain a rough estimate on the expected PDR fluxes in different lines, we have evaluated the PDR fluxes that we would expect for a system like the Seyfert 2 galaxy NGC~1068, if it were placed at high redshift. We adopt $z=8$. This system consists of a central X-ray dominated region \citep{Galliano03} and a circumnuclear starburst ring of $\sim3$~kpc in size, with a stellar mass of $\sim10^6\ M_\odot$ and an age of $5$~Myr \citep{Spinoglio05}. On scales of a few hundred parsecs, one finds a star formation rate of a few times $10\ M_\odot$~yr$^{-1}$~kpc$^{-2}$ \citep{Davies07}. This is close to the star formation rate in Eddington-limited starbursts as suggested by \citet{Thompson05}. To understand which of the fluxes emitted in this region would be detectable with ALMA if this system were located at $z=8$, we went through the spectroscopic sample of \citet{Spinoglio05} and checked which lines would be redshifted into the ALMA frequency bands, and what would be the expected amount of flux. The fluxes given by \citet{Spinoglio05} have been measured with a frequency resolution of $1500$~km/s. Correspondingly high velocities can indeed be reached in the presence of fast jets or outflows. However, typical line profiles show that most of the flux is in a range of $\pm150$~km/s, which we adopt here as a fiducial value. The so obtained observable line transitions are summarized in Table~\ref{tab:observables}, while the expected sensitivity and angular resolution at the largest baselines is given in Table~\ref{tab:ALMA}. For all the transitions, a $3\sigma$ detection seems possible for an integration time of a few hours. The table does not include CO transitions, as the PDR would only excite the low-lying transitions which would not fall in ALMA's frequency range for $z>8$. {We note that the expected ratio between [NII] and [CII] is comparable to the observational upper limit derived by \citet{Walter09b}.}

\section{Expectations for the XDR}\label{XDR}
{As ALMA may for the first time detect and resolve emission for the central X-ray dominated regions, we want to assess here in more detail the expected chemical conditions in the central region and the corresponding fluxes in different lines. We start by discussing the implications of X-rays for the conditions in molecular clouds. We then show how ALMA observations can distinguish between X-ray chemistry and an intense burst of star formation on the same spatial scales. Afterwards, we provide a set of systematic model predictions, first assuming an XDR of constant size, but also considering the potential increase of the XDR in case of a higher X-ray luminosity.}

\subsection{Implications of X-rays for molecular clouds}
{We follow the chemistry in a one-dimensional molecular cloud complex irradiated by X-rays with the XDR code of \citet{Meijerink05}. } The model includes more than $50$ chemical species and several thousand reactions. For CO, the detailed level populations are solved consistently with the 1D radiation transport equation \citep{Poelman05, Poelman06}. As the low-metallicity case was explored in detail by \citet{Spaans08}, we focus here in particular on situations with about solar metallicity. 

\begin{figure}[t]
\includegraphics[scale=0.55]{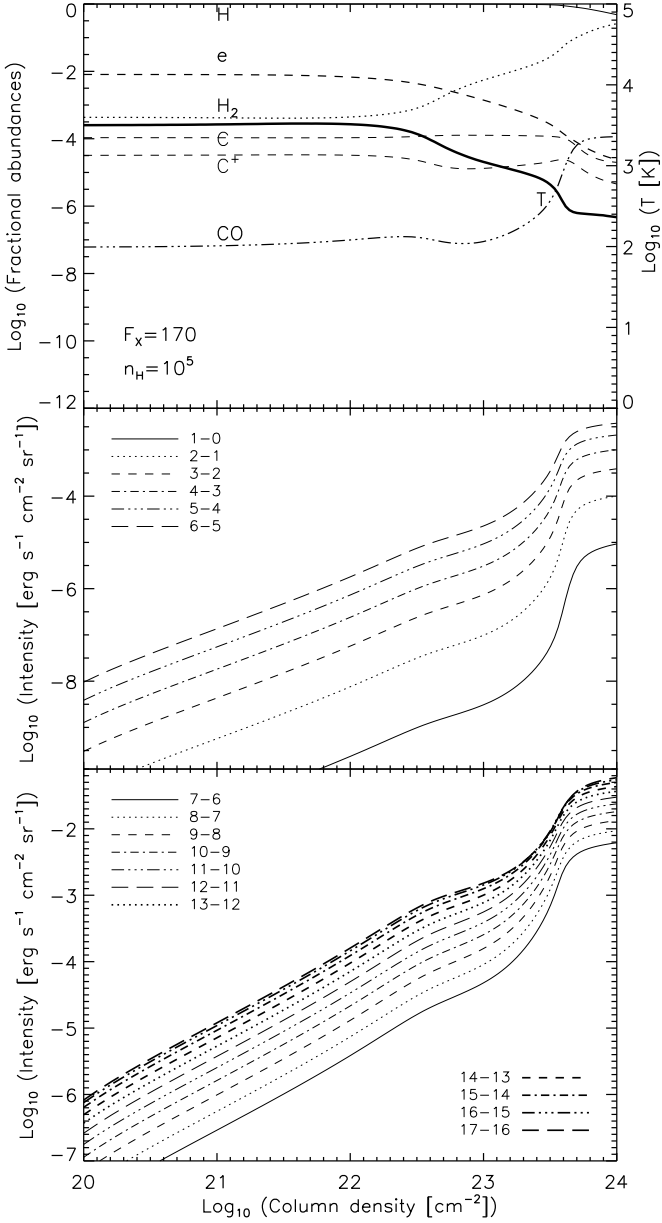}
\caption{A model for the X-ray chemistry in NGC~1068. The adopted flux impinging on the cloud is $170$~erg~s$^{-1}$~cm$^{-2}$. The adopted density is $10^5\ $cm$^{-3}$. Top: The abundances of different species as a function of column density. Middle: The low-$J$~CO lines as a function of column density. Bottom: The high-$J$~CO lines as a function of column density. }
\label{fig:ngc1068}
\end{figure}

The first model we discuss corresponds to the XDR in the Seyfert 2 galaxy NGC~1068 (see \S~\ref{ngc1068}). We plot chemical abundances and CO emission for cloud column densities between $10^{20}-10^{24}$~cm$^{-2}$ in Fig.~\ref{fig:ngc1068}. Larger column densities correspond to extreme ULIRGs like Arp~220 that is even optically thick around $350$~GHz (P. Papadopoulos, private communication). The fiducial gas density of $10^5$~cm$^{-3}$ has little impact on our results, unless it drops to below $10^{4.5}$~cm$^{-3}$. Above this limit, the XDR properties are determined by the ratio of X-ray flux to gas density. For lower densities, emission in the high-$J$ CO lines would not be excited due to the critical densities. However, high-density gas appears to exist in the center of NGC~1068 \citep{Galliano03}.

\begin{figure}[t]
\includegraphics[scale=0.55]{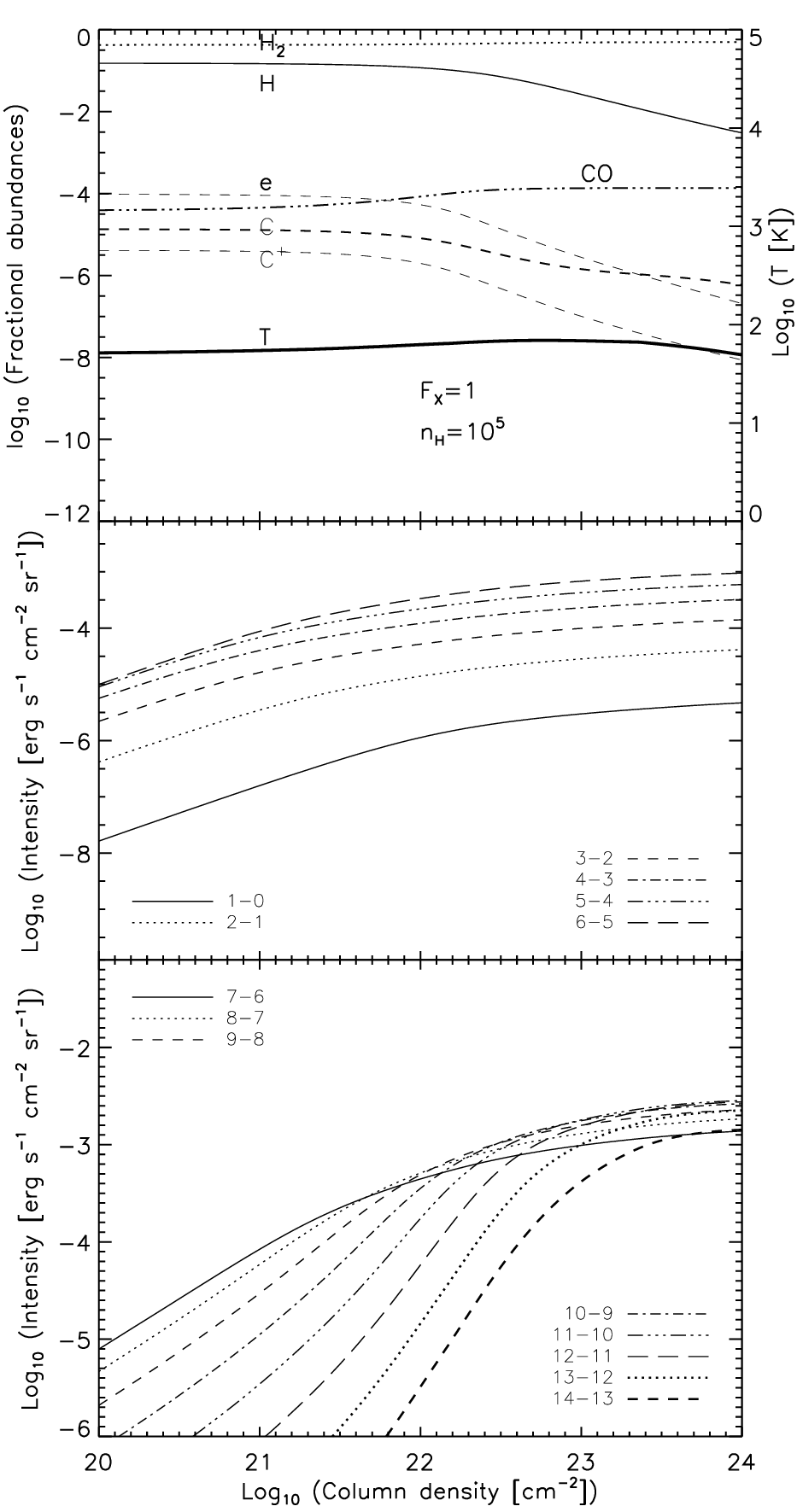}
\caption{The X-ray chemistry in a system with X-ray flux of $1$~erg~s$^{-1}$~cm$^{-2}$ impinging on the cloud. The adopted density is $10^5\ $cm$^{-3}$. Top: The abundances of different species as a function of column density. Middle: The low-$J$~CO lines as a function of column density. Bottom: The high-$J$~CO lines as a function of column density. }
\label{fig:ngc1}
\end{figure}

\begin{figure}[t]
\includegraphics[scale=0.55]{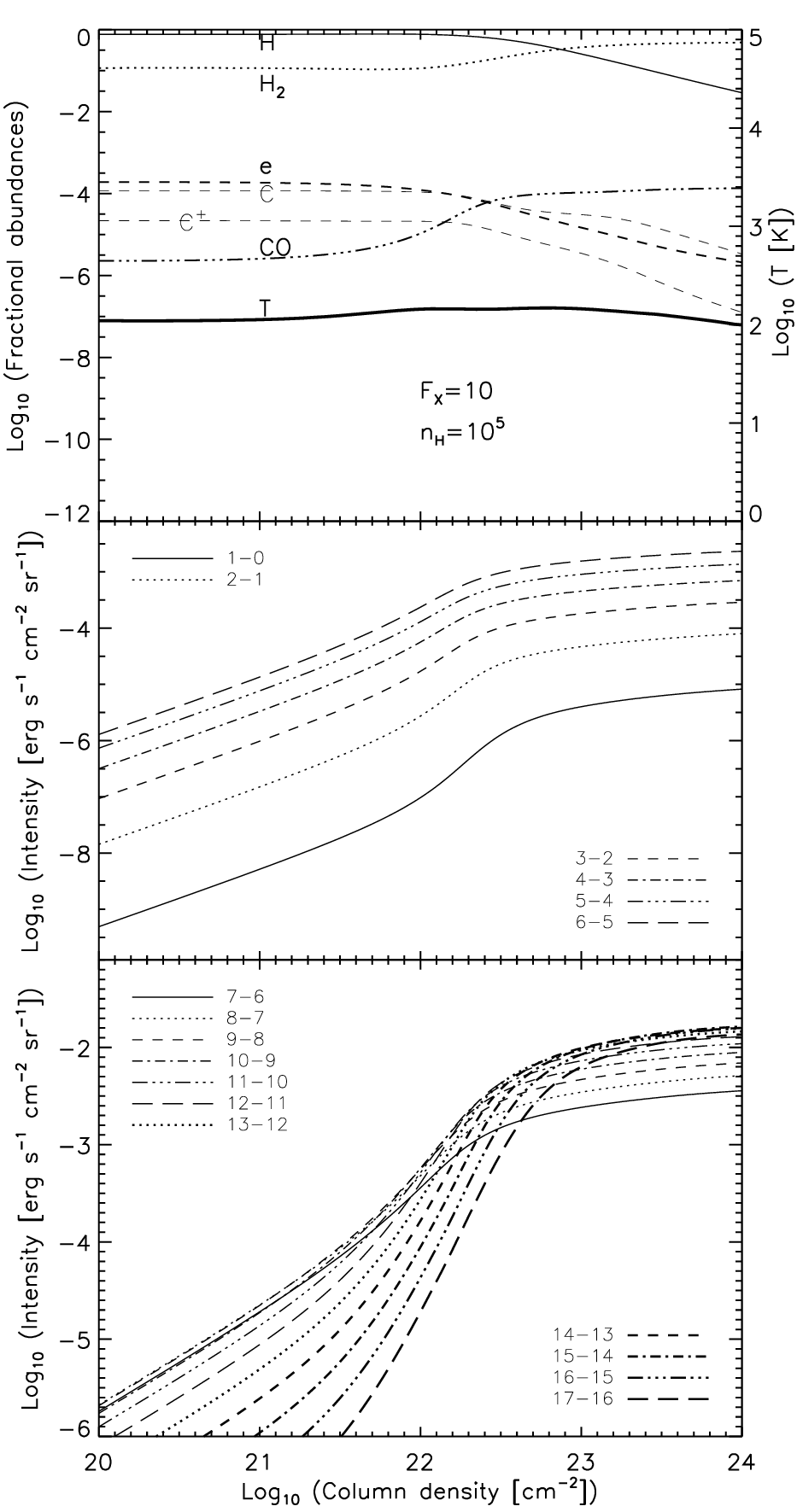}
\caption{The X-ray chemistry in a system with X-ray flux of $10$~erg~s$^{-1}$~cm$^{-2}$ impinging on the cloud. The adopted density is $10^5\ $cm$^{-3}$. Top: The abundances of different species as a function of column density. Middle: The low-$J$~CO lines as a function of column density. Bottom: The high-$J$~CO lines as a function of column density. }
\label{fig:ngc10}
\end{figure}

The strong X-ray flux of $\sim170$~erg~s$^{-1}$~cm$^{-2}$ in NGC~1068 suffices to make the gas essentially atomic and leads to high temperatures of $\sim3000$~K, as well as relatively low CO abundances of the order $10^{-7}$ due to photodissociation by soft UV-photons produced after the absorption of X-rays. However, the CO abundance is still higher than in typical PDRs, and the CO intensity is high, due to the strong thermal excitation in the hot gas. For a column of $10^{22}\ \mathrm{cm}^{-2}$, our results appear of the same magnitude as in the model of \citet{Galliano03}. For larger columns, the temperature gradually decreases, the gas becomes molecular and CO gets more abundant, and we find intensities of the order $10^{-2}$~erg~s$^{-1}$~cm$^{-2}$~sr$^{-1}$ in the high-$J$~CO lines.

To explore the dependence of the chemistry on the X-ray flux, we consider two additional cases.  An extreme case with $\sim1$~erg~s$^{-1}$~cm$^{-2}$ is shown in Fig.~\ref{fig:ngc1}. In this model, we find lower temperatures of $\sim70$~K, a large fraction of molecular gas and CO abundances of the order of $10^{-4}$. While the lower temperature tends to decrease the CO line intensities, they are still enhanced due to the larger CO abundance. Above a column of $10^{23}\ \mathrm{cm}^{-2}$, the intensities increase rather slowly as the lines become optically thick. 

As an intermediate scenario, we consider a source with an X-ray flux of $\sim10$~erg~s$^{-1}$~cm$^{-2}$ (see Fig.~\ref{fig:ngc10}). In this model, the temperature is increased to $\sim100$~K. The CO abundance is initially of the order $3\times10^{-6}$ and increases to $\sim10^{-4}$ for larger columns. For columns less than $10^{22}\ \mathrm{cm}^{-2}$, the intensities are thus reduced by about an order of magnitude compared to the previous case, while they are increased by an order of magnitude for larger columns. 

\subsection{Separating the XDR from a nuclear starburst}
\begin{figure}[t]
\includegraphics[scale=0.55]{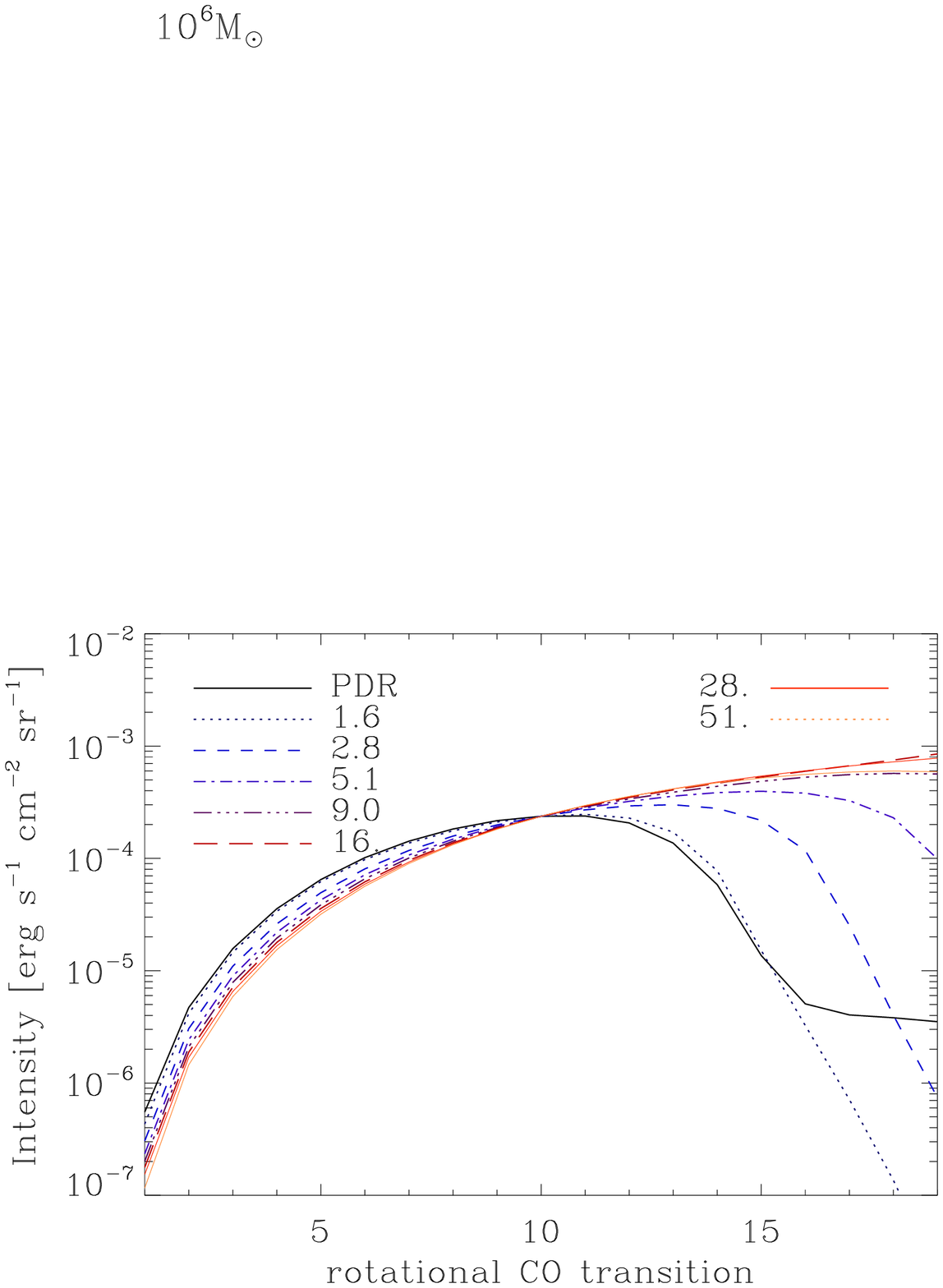}
\caption{A comparison of the CO line SED in case of an intense starburst with $G_0=10^5$ with the corresponding SED for molecular clouds under X-ray irradiation, for different X-ray fluxes in erg~s$^{-1}$~cm$^{-2}$. The spectra are normalized such that they have the same intensity in the 10th transition. If the impinging X-ray flux is at least 2.8~erg~s$^{-1}$~cm$^{-2}$, observations of the 15th CO rotational transition can clearly discriminate between PDR and XDR chemistry. }
\label{fig:spectra}
\end{figure}

{
In the center of an active galaxy, not only the X-ray emission is enhanced, but one may expect the presence of a strong nuclear starburst. For instance, Arp~$220$ harbors such a starburst on scales of $\sim300$~pc. We therefore compare the expected CO line SED of a strong starburst with $G_0=10^5$ with the CO line SEDs in X-ray dominated regions, based on the models provided by \citet{Meijerink07}. We normalize them such that the CO (10-9) transition has the same intensity in all models. In this case, the SEDs can hardly be distinguished at the low-$J$ transitions that are typically observed at low redshift (see Fig.~\ref{fig:spectra}). At higher-$J$ transitions, the PDR SED drops considerably and flattens on a low level due to the small amount of hot gas in the outer layer of the molecular cloud. We expect that a value of $G_0=10^5$ is a robust upper limit for the soft-UV flux that can be obtained in a galaxy. In fact, larger values have never been indicated in previous observations, and indeed such a value would require extreme conditions as in the Orion Bar throughout all of the galaxy. In XDRs, a much larger fraction of the gas is at high temperatures, and thus the SED is not expected to drop as rapidly. }

{
If the total amount of energy injected by X-rays and by soft-UV photons is comparable (within a factor of 10), then the presence of X-rays can be clearly inferred using the CO (16-15) transition, if the local X-ray flux is at least $2.8$~erg~s$^{-1}$~cm$^{-2}$. However, as discussed in \S~\ref{heat}, we expect X-ray flux to dominate over the soft-UV in the center of the galaxy. Observations at even higher-$J$ transitions may be useful to determine the local amount of X-ray flux from the CO line SED. A potential uncertainty is the presence of cold dust, which may to some degree absorb the CO line emission and thus change the appearance of the SED. Due to the characteristic scaling of dust absorption with wavelength, we expect that such a behavior could be recognized and potentially corrected. For this purpose, it is of course desirable to measure as many high-$J$ CO lines as possible.
}

\subsection{Model predictions for XDRs of constant size}
{
Although we have shown in \S~\ref{heat} that the central XDRs can likely be resolved with ALMA, it is currently not clear how the expected XDR size varies with X-ray luminosity. If the strengths of the soft-UV field is independent of this, one should on average expect a larger XDR for higher X-ray luminosities. However, it is also conceivable that the X-ray luminosity is indicative of the system as a whole, and that a higher X-ray luminosity may be accompanied by a stronger soft-UV field. In such a case, one might expect a smaller increase in the XDR size or even a constant size. For this reason, we will consider two extreme cases, assuming that more realistic scenarios should lie in between the two. In this subsection, we will assume that the size of the XDR is always constant, of $\sim200$~pc. Of course, the numbers given here can be easily rescaled for other XDR sizes, or for area-filling factors smaller than 1. In the following subsection, we will then discuss the implications of varying the X-ray luminosity for constant $G_0$.}

\begin{figure}[t]
\includegraphics[scale=0.5]{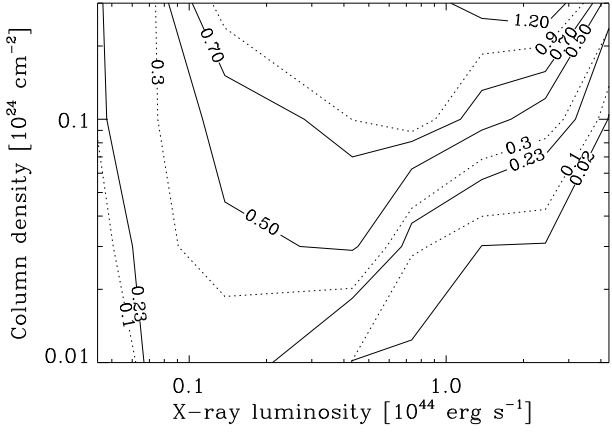}
\caption{The expected flux in mJy for high-$J$ CO lines, for a central XDR of $200$~pc, and molecular clouds of $10^5$~cm$^{-3}$, as a function of X-ray luminosity and cloud column density. We focus on lines that fall in ALMA band 6, which offers a good compromise between angular resolution and sensitivity. For a source at $z=5$ (solid line), this corresponds to the (10-9) CO transition, for a source at $z=8$, it corresponds to the (14-13) CO transition.}
\label{fig:CO}
\end{figure}

\begin{figure}[t]
\includegraphics[scale=0.5]{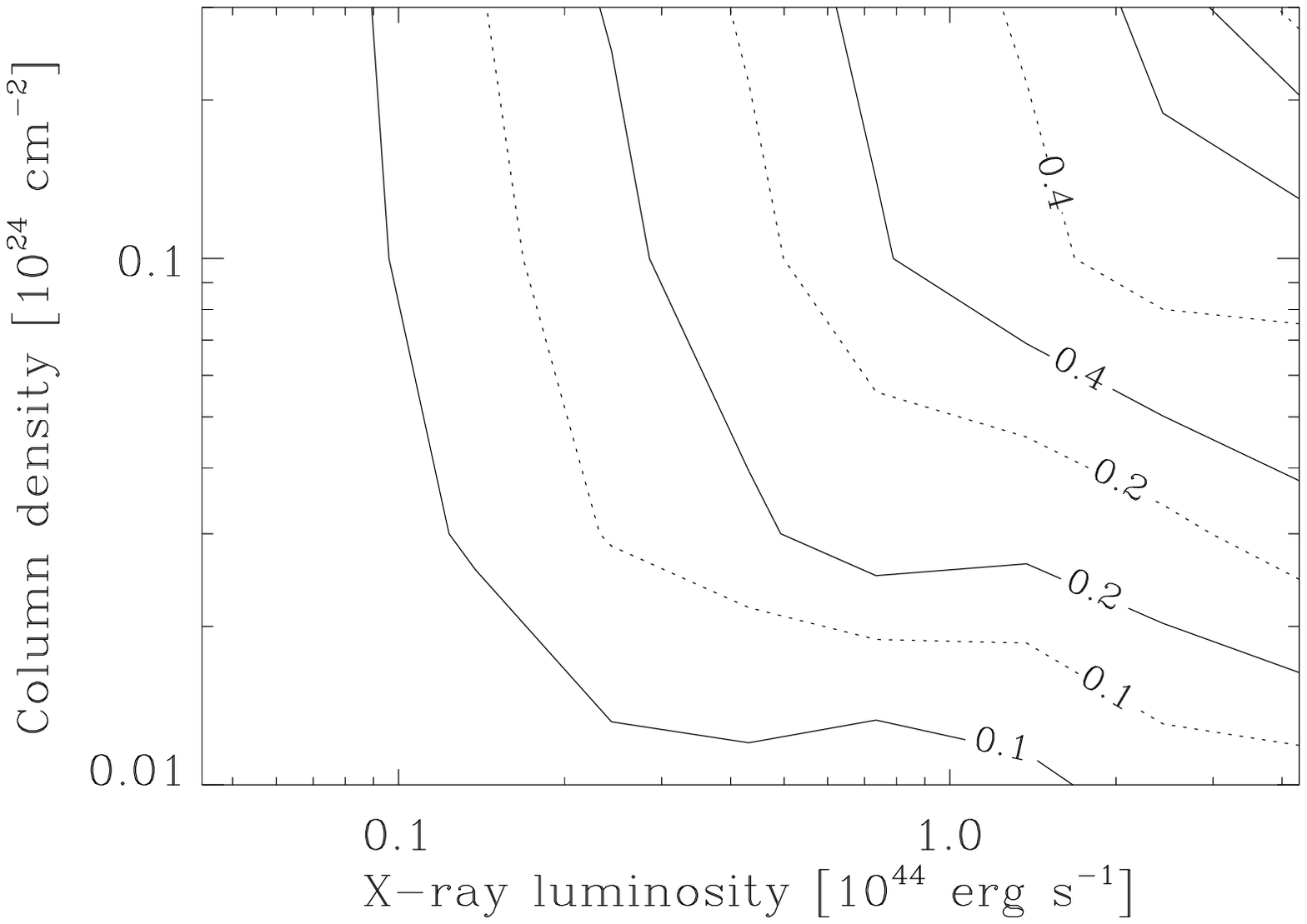}
\caption{The expected flux in mJy for the [CII]~$158$~$\mu$m line, for a central XDR of $200$~pc, and molecular clouds of $10^5$~cm$^{-3}$, as a function of X-ray luminosity and cloud column density. The solid line corresponds to a source at $z=5$, and the dashed line to a source at $z=8$. At $z=5$, the line is redshifted into ALMA band $7$, with a sensitivity of $0.08$~mJy (1 day, $3\sigma$, $300$~km/s). At $z=8$, it falls into ALMA band 6 with a sensitivity of $0.04$~mJy.}
\label{fig:CII}
\end{figure}

\begin{figure}[t]
\includegraphics[scale=0.5]{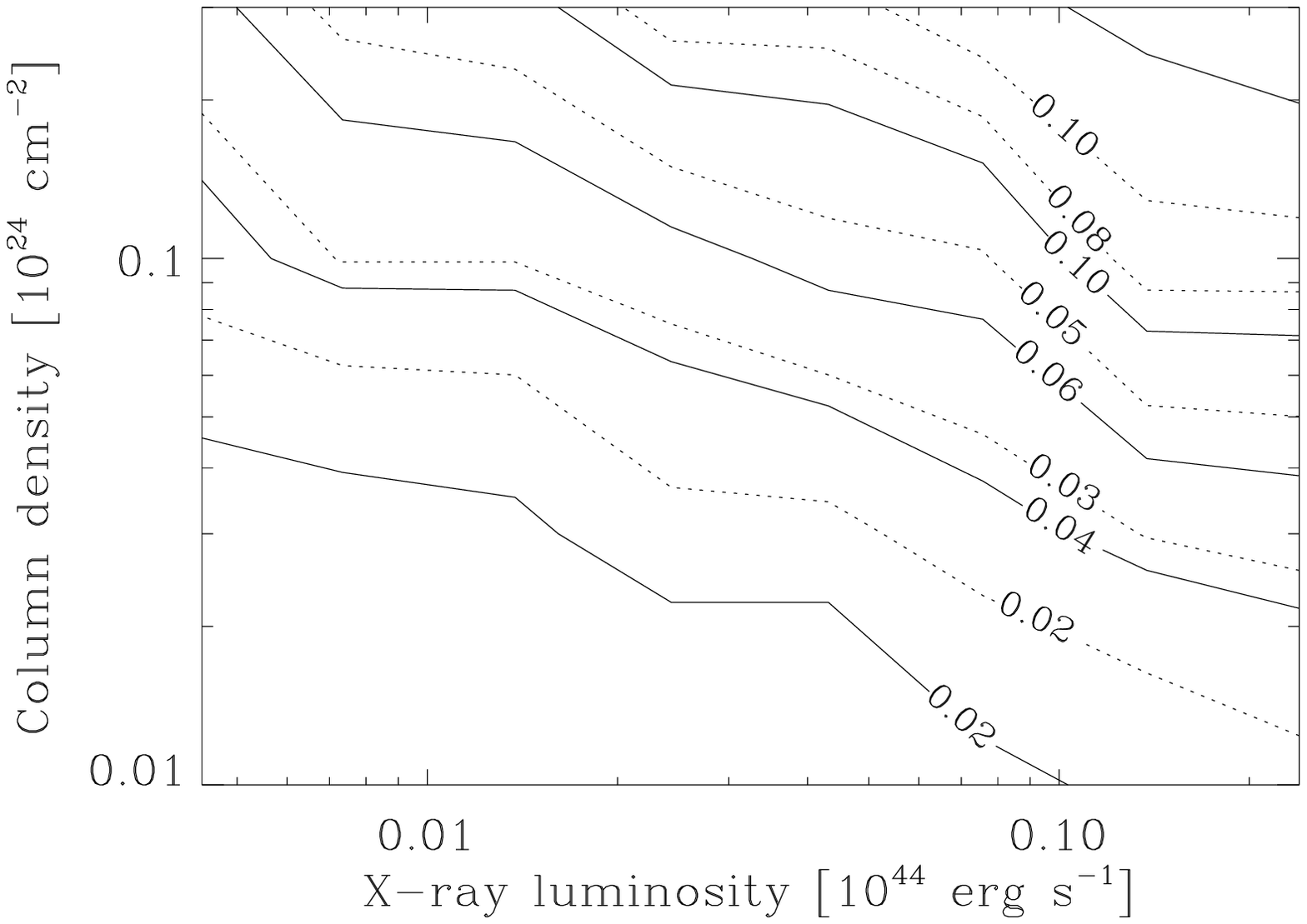}
\caption{The expected flux in mJy for the [CII]~$158$~$\mu$m line, for a central XDR of $200$~pc, and molecular clouds of $10^4$~cm$^{-3}$, as a function of X-ray luminosity and cloud column density. The solid line corresponds to a source at $z=5$, and the dashed line to a source at $z=8$. At $z=5$, the line is redshifted into ALMA band $7$, with a sensitivity of $0.08$~mJy (1 day, $3\sigma$, $300$~km/s). At $z=8$, it falls into ALMA band 6 with a sensitivity of $0.04$~mJy.}
\label{fig:CII1e4}
\end{figure}

\begin{figure}[t]
\includegraphics[scale=0.5]{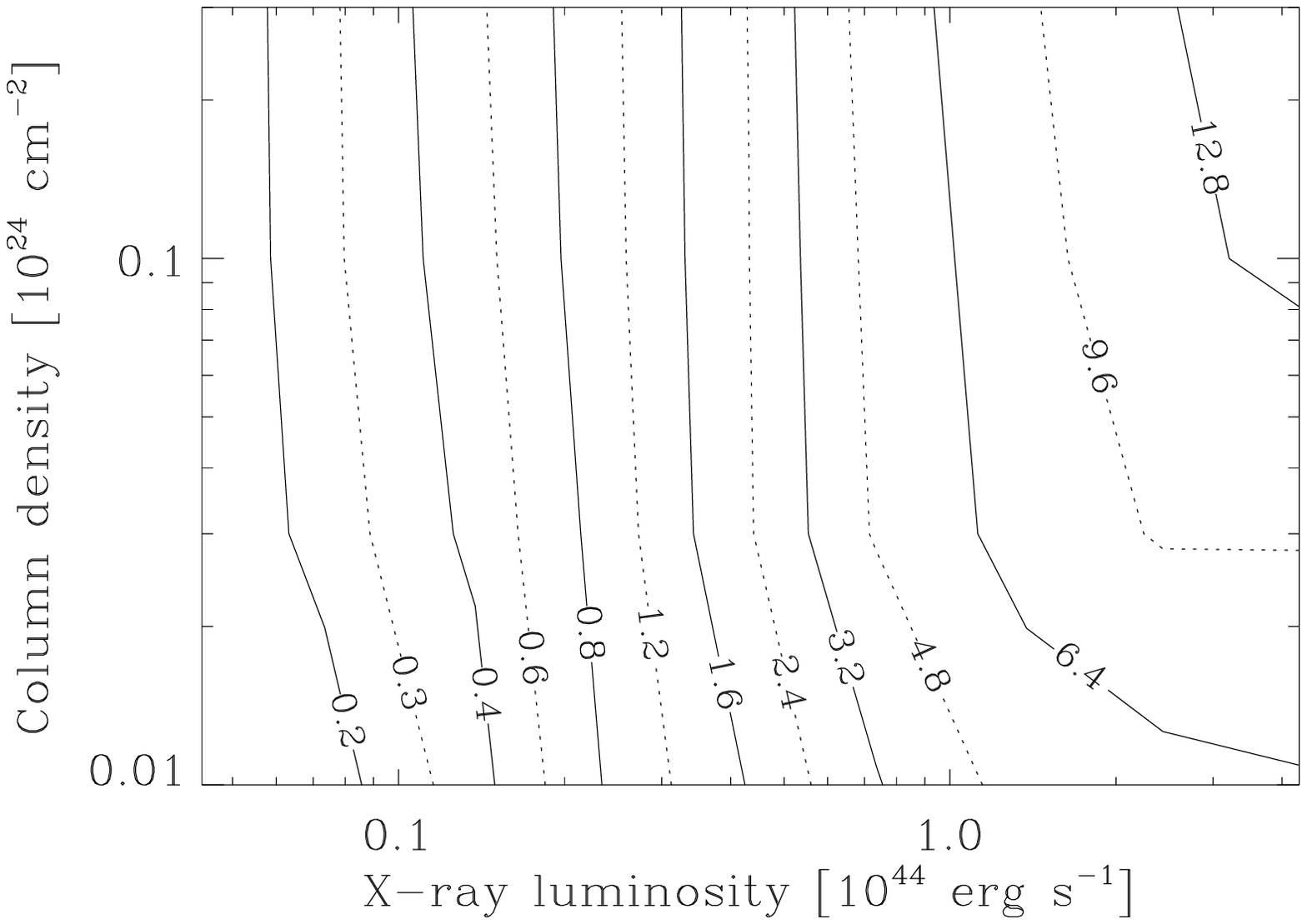}
\caption{The expected flux in mJy for the [OI]~$63$~$\mu$m line, for a central XDR of $200$~pc, and molecular clouds of $10^5$~cm$^{-3}$, as a function of X-ray luminosity and cloud column density. The solid line corresponds to a source at $z=5$, and the dashed line to a source at $z=8$. At $z=5$, the line is redshifted into ALMA band $10$, with a sensitivity of $0.2$~mJy (1 day, $3\sigma$, $300$~km/s). At $z=8$, it falls into ALMA band 8 with a sensitivity of $0.13$~mJy.}
\label{fig:OI63}
\end{figure}

\begin{figure}[t]
\includegraphics[scale=0.5]{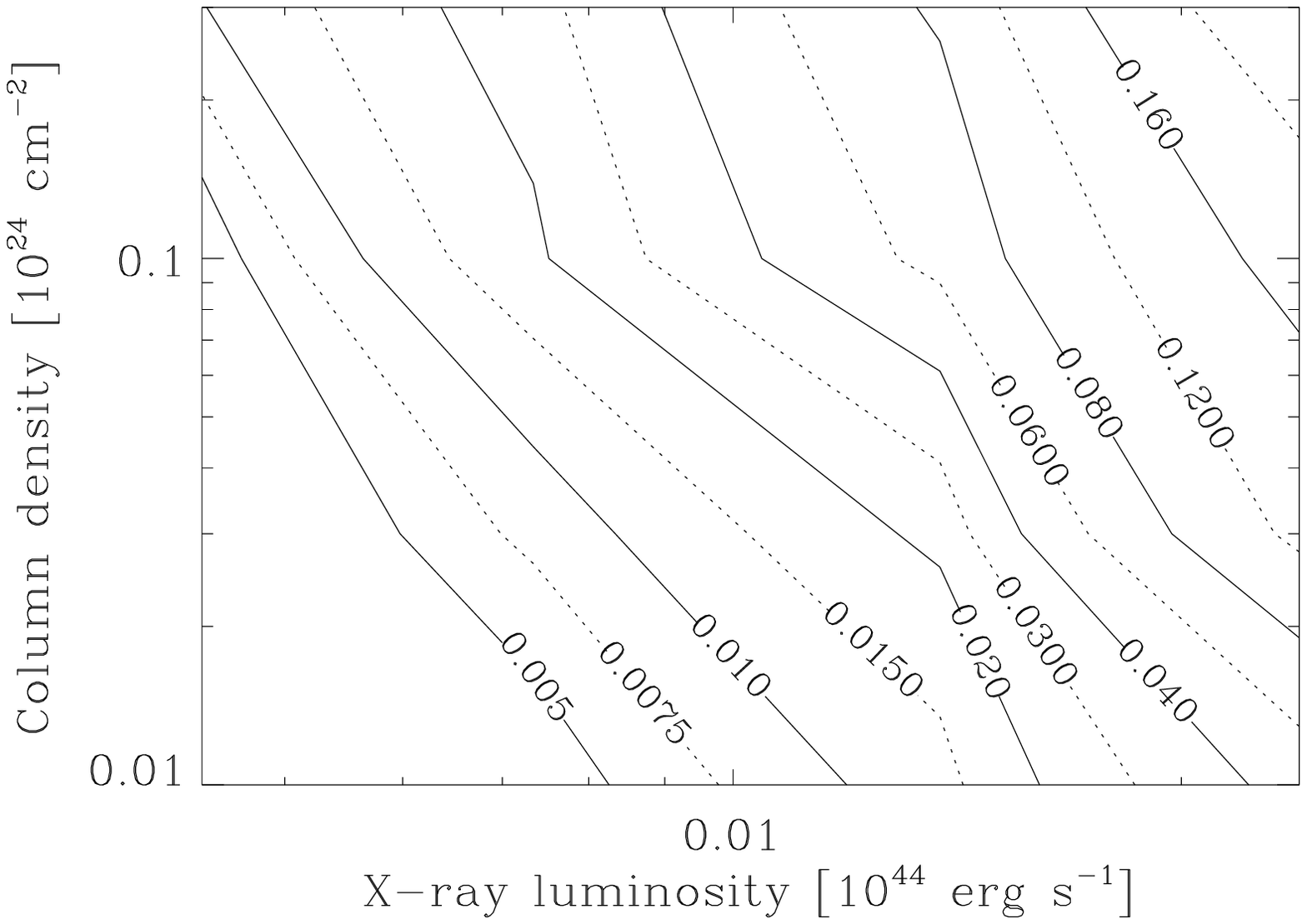}
\caption{The expected flux in mJy for the [OI]~$63$~$\mu$m line, for a central XDR of $200$~pc, and molecular clouds of $10^4$~cm$^{-3}$, as a function of X-ray luminosity and cloud column density. The solid line corresponds to a source at $z=5$, and the dashed line to a source at $z=8$. At $z=5$, the line is redshifted into ALMA band $10$, with a sensitivity of $0.2$~mJy (1 day, $3\sigma$, $300$~km/s). At $z=8$, it falls into ALMA band 8 with a sensitivity of $0.13$~mJy.}
\label{fig:OI631e4}
\end{figure}

\begin{figure}[t]
\includegraphics[scale=0.5]{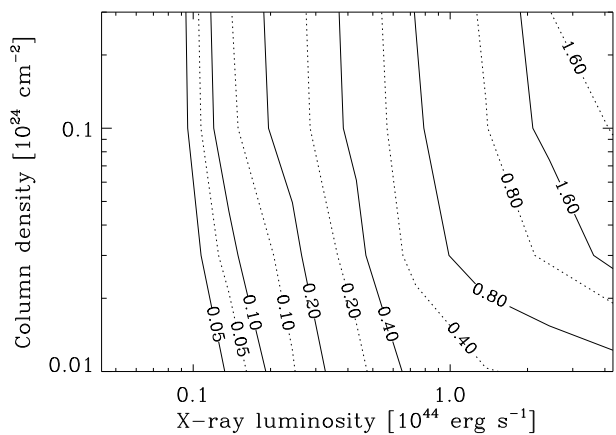}
\caption{The expected flux in mJy for the [OI]~$146$~$\mu$m line, for a central XDR of $200$~pc, and molecular clouds of $10^5$~cm$^{-3}$, as a function of X-ray luminosity and cloud column density. The solid line corresponds to a source at $z=5$, and the dashed line to a source at $z=8$. At $z=5$, the line is redshifted into ALMA band $7$, with a sensitivity of $0.37$~mJy (1 day, $3\sigma$, $300$~km/s). At $z=8$, it falls into ALMA band 6 with a sensitivity of $0.225$~mJy.}
\label{fig:OI146}
\end{figure}

For {a first estimate}, let us consider at source at $z=5$ with an XDR of at least $100$~pc, corresponding to an angular scale of $0.016$'', and a typical intensity in the high-$J$~CO lines of $10^{-3}$~erg~s$^{-1}$~cm$^{-2}$~sr$^{-1}$. As can be seen from the calculations above, such an intensity can be reached in a broad range of systems for column densities of at least $10^{23}$~cm$^{-2}$, and in fact also for column densities of at least $10^{22}$~cm$^{-2}$ in the presence of sufficient X-ray flux. With a fiducial velocity dispersion of $300$~km/s, this corresponds to a flux of $0.03$~mJy, which is detectable in a bit more than a day in ALMA band 6. 

In a similar way, it is possible to calculate the expected fluxes also in various fine-structure lines. Based on the detailed parameter study provided by \citet{Meijerink07}\footnote{http://www.strw.leidenuniv.nl/~meijerin/grid/}, which shows the expected fluxes in various lines as a function of X-ray luminosity density and column density, we therefore provide detailed predictions for the fluxes from the central X-ray dominated region, assuming a characteristic size of $200$~pc, consistent with the results obtained in \S~\ref{heat}. 

For the high-$J$ CO lines, we focus on those which are redshifted into ALMA band 6, which offers a good compromise between angular resolution ($0.014$'' at the largest baseline) and sensitivity ($\sim0.04$~mJy for an integration time of one day, $3\sigma$ detection and a line width of $300$~km/s). At a redshift $z=5$, this corresponds to the (10-9) CO transition, at $z=8$ to the (14-13) CO transition. We assume that the projected surface area is homogeneously filled with molecular clouds with central densities of $10^5$~cm$^{-3}$. Fig.~\ref{fig:CO} shows how the expected flux varies as a function of the cloud column density and the X-ray luminosity\footnote{For the conversion between X-ray luminosity and flux, we assumed optically thin conditions. One may need to correct for further attenuation in case of large filling-factors.}. Similar results would be obtained for cloud densities of $10^{4.5}$~cm$^{-3}$, while it is more difficult to excite CO emission at lower densities. 

The figure illustrates that higher fluxes can be obtained for larger column densities, while intermediate X-ray fluxes are ideal for stimulating emission in the CO lines considered here. This is because at very high fluxes, a significant amount of X-rays would be converted into soft UV-photons and dissociate the molecules.

In Figs.~\ref{fig:CII} and \ref{fig:CII1e4}, we show the corresponding results for the [CII]~$158$~$\mu$m line, both for a density of $10^5$~cm$^{-3}$ and a density of $10^4$~cm$^{-3}$. For densities of $10^5$~cm$^{-3}$, low column densities are sufficient to yield a detectable amount of flux, and the flux monotonically increases with the X-ray luminosity. As in the PDR case, this line is therefore valuable to explore the centers of high-redshift quasars, and provides complementary information to the CO lines. For densities of $10^4$~cm$^{-3}$, we find a stronger dependence on column density, and indeed columns of at least $10^{23}$~cm$^{-2}$ are needed to yield a detectable amount of flux.

The results for the [OI]~$63$~$\mu$m line are given in Figs.~\ref{fig:OI63} and \ref{fig:OI631e4}, again for densities of $10^5$~cm$^{-3}$ and $10^4$~cm$^{-3}$, respectively. This line quickly becomes optically thick. Therefore, in particular for cloud densities of $10^5$~cm$^{-3}$, it is insensitive to the column density, but provides a good measure for the X-ray flux. It is very bright. Even for cloud densities of $10^4$~cm$^{-3}$, it depends more on X-ray luminosity than column density, and is still detectable in the case of high column densities and strong X-ray fluxes. It is thus well-suited to study gas dynamics in the central XDR by resolving the line profile.

We also provide results for the [OI]~$146$~$\mu$m line in Fig.~\ref{fig:OI146}, for a cloud density of $10^5$~cm$^{-3}$. For lower densities, this line is hard to excite. It can also be very bright and shows a strong dependence on the X-ray flux. {The ratio between the [OI] $63$~$\mu$m line and the $145$~$\mu$m line is generally about $0.1$.}

Emission from neutral carbon seems more difficult to detect. The intensity of the [CI]~$369$~$\mu$m line is typically {at least} an order of magnitude smaller than the intensity in the [CII]~$158$~$\mu$m line, and only significant in the case of strong X-ray fluxes and high column densities. {As discussed in \S~\ref{quasar6}, the relatively low ratio between these lines in the recently detected $z=6.42$ quasar puts a significant constraint on theoretical models.} Other carbon lines like [CI]~$609$~$\mu$m or [CI]~$230$~$\mu$m are even weaker and should never be visible. Additional lines like [SiII]~$35$~$\mu$m or [FeII]~$26$~$\mu$m can be bright as well \citep{Meijerink07}, but typically have too short wavelengths for ALMA, except at redshifts $z>9$.

{The reader may notice that for the model predictions given in this subsection, we restricted ourselves to a range of X-ray luminosities of about two orders of magnitude, corresponding to the range of X-ray fluxes avaiable in the data by \citet{Meijerink07}. This range of data has been chosen such that it covers the observationally interesting cases. For lower fluxes, or in our case lower X-ray luminosities, we would not expect significant emission driven by X-rays, as can be seen in the corresponding figures. Similarly, it is straightforward to extrapolate the behavior towards higher X-ray fluxes: As shown in Fig.~\ref{fig:CO}, the CO intensities decrease considerably at higher luminosities. This is because part of the X-rays are reprocessed to soft-UV photons that dissociate all the CO. At even higher intensities, the gas temperature would increase up to $10^4$~K and become fully ionized. Molecules would no longer survive under such circumstances. The emission of [CII] may increase a bit further, until a large fraction of carbon is doubly-ionized at temperatures near $10^4$~K. Similar considerations hold for the oxygen line. However, it is not clear whether such scenarios are actually physical, or if the molecular cloud would rather be photo-evaporated at that point. As already mentioned above, the size of the XDR may be more extended in the case of such high luminosities. Then, moderate X-ray fluxes will be present on larger scales. }

In summary, we can therefore conclude that the main observables for the central XDR are the high-$J$ CO lines as well as the fine-structure lines of [CII] and [OI]. The [OI]~$63$ and $146$~$\mu$m lines show a strong dependence on X-ray luminosity and may provide a good handle on this quantity. In combination with an observation of the [CII] or high-$J$ CO lines, the X-ray luminosity can be estimated as well. As shown by \citet{Meijerink07}, such line ratios also provide valuable information on gas density and temperature.  We also note that it is possible to discriminate such XDRs from regions with strong mechanical heating \citep{Papadopoulos08}. This can be done for instance on the basis of the observed X-ray luminosity, or by looking at the dust SED. While in XDRs, both dust and gas will be at high temperatures in dense clouds, there should be a clear discrepancy between gas and dust temperature if heating is due to local shocks. We therefore expect that the physical conditions in the central XDRs can be probed in detail with ALMA. 

\subsection{Model predictions for variable XDR-sizes}

\begin{figure}[t]
\includegraphics[scale=0.5]{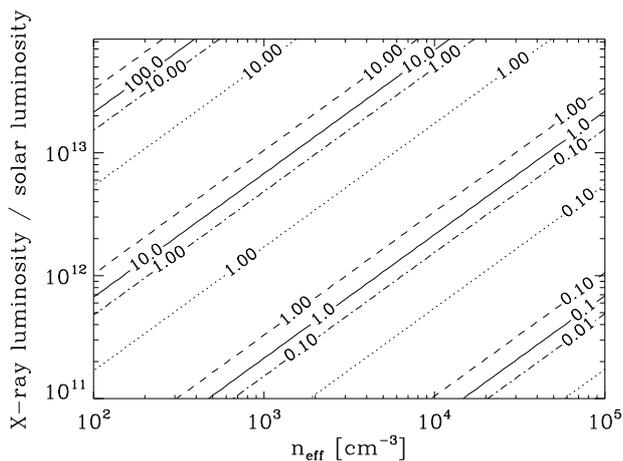}
\caption{The expected flux in mJy for the CO (14-13) transition (solid line), [CII] $158$~$\mu$m (dotted line), [OI]~$63$~$\mu$m (dashed line) and [OI]~$146$~$\mu$m (dot-dashed line) emission for a quasar at $z=5$, as a function of X-ray luminosity and average density. We assume a typical cloud column density of $10^{23}$~cm$^{-2}$, and a soft-UV field $G_0=100$.}
\label{fig:modellum}
\end{figure}

\begin{figure}[t]
\includegraphics[scale=0.5]{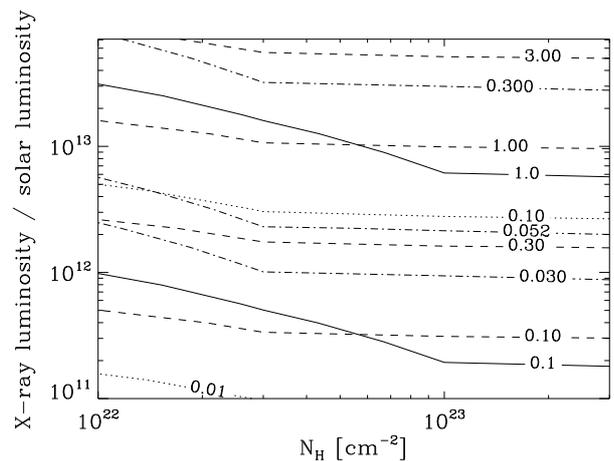}
\caption{The expected flux in mJy for the CO (14-13) transition (solid line), [CII] $158$~$\mu$m (dotted line), [OI] $63$~$\mu$m (dashed line) and [OI]~$146$~$\mu$m (dot-dashed line) emission for a quasar at $z=8$, as a function of X-ray luminosity and cloud column density. We assume a typical cloud density of $10^{4}$~cm$^{-3}$, and a soft-UV field $G_0=100$.}
\label{fig:modelcol}
\end{figure}

{As mentioned above, the size of the XDR may increase considerably for increasing X-ray luminosity. We first explore the dependence on the average density and the X-ray luminosity, assuming a typical cloud column density of $10^{23}$~cm$^{-2}$, and a soft-UV field $G_0=100$. The X-ray spectrum is assumed to range from $1$ to $100$~keV. As shown in \S~\ref{heat}, the size of the XDR then varies as a power-law with density. The same is true for the expected fluxes of CO, [CII] and [OI], as shown in Fig.~\ref{fig:modellum}. We estimate those by adopting a typical value in mJy on scales corresponding to the size of the XDR. These should indeed consitute the main contribution. For [CII] and [OI], the total flux may be a bit larger, as the expected amount of emission increases with the X-ray flux (see previous subsection). CO, on the other hand, might be efficiently dissociated in the very inner core. However, we do not expect this to affect our predictions significantly.
}

{We also explore the role of the molecular cloud column density for the expected fluxes. For this purpose, we adopt an average density of $10^4$~cm$^{-3}$ and a soft-UV field $G_0=100$. In Fig.~\ref{fig:modelcol}, we show the expected fluxes as a function of X-ray luminosity and column density. As a generic feature, we find that the expected fluxes vary for columns smaller than $10^{23}$~cm$^{-2}$, but level-off for higher values. The fluxes in high-column density systems should thus essentially depend on the X-ray luminosity only.}

\section{Evidence for XDRs and the interpretation of sub-mm line observations}\label{evidence}
Although the presence of a central X-ray dominated region seems unavoidable from a theoretical point of view, we want to review current evidence for the presence of central XDRs in active galaxies. Such evidence is present in local galaxies like NGC~1068 that can be studied in great detail, as well as in high-redshift quasars like APM~08279 in which the corresponding fluxes are magnified by a gravitational lens. Similar indications are present also in the Cloverleaf quasar, where CO fluxes up to the (9-8) transition have been measured, and no turnover in the line SED has been found yet \citep{Bradford09}. These sources will allow a first test for the expectations we have formulated in the previous section once the ALMA telescope becomes available. {We also discuss the $z=6.42$ quasar SDSS~J114816.64+525150.3, as it provides a good and interesting example concerning the complex interpretation of sub-mm line observations. We will further discuss how its properties can be understood better if additional data are provided, with particular focus on the role of ALMA.}

To distinguish between different excitation mechanisms of molecular clouds, it is very important to have observational diagnostics for the various excitation mechanisms. First efforts for modelling the chemistry in XDRs were performed by \citet{Maloney96} and \citet{Lepp96}, while PDR chemistry was originally studied by \citet{Hollenbach99}. Recent efforts to discriminate such models have been performed by \citet{Perez07, Perez09}, and new diagnostic diagrams to discriminate AGN- and starburst-dominated galaxies have been provided by \citet{Spoon07} and \citet{Hao09}.

\subsection{NGC 1068}\label{ngc1068}

\begin{figure}[t]
\includegraphics[scale=0.4]{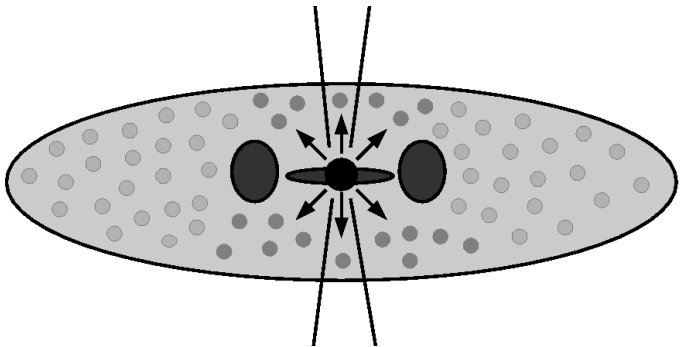}
\caption{A sketch for a situation with an inhomogeneous XDR, motivated by the observations of \citet{Galliano03} in NGC~1068. While X-rays are shielded by a central absorber along the line of sight, they may stimulate emission in molecular clouds in the perpendicular direction with the typical characteristics of XDRs.}
\label{fig:quasar}
\end{figure}

As shown by \citet{Galliano03}, NGC~1068 contains a central XDR. Their analysis is based on the observed intensitiy of the H$_2$~$2.12$~$\mu$m line \citep{Galliano02} and the rotational CO lines \citep{Schinnerer00}, and showed that the observed emission can be consistently explained with the XDR-model of \citet{Maloney96} under the following assumptions:

\begin{itemize}
\item The central engine is a power-law X-ray source with spectral slope $\alpha=-0.7$ and luminosity of $10^{44}$~erg~s$^{-1}$ in the $1-100$~keV range, consistent with the X-ray luminosity determined from VLBI water maser observations \citep{Greenhill96}.
\item The emission originates from molecular clouds with a density of $10^5$~cm$^{-3}$, a column of $10^{22}$~cm$^{-2}$ at a distance of $70$~pc and solar metallicity.
\end{itemize}

They show that the central XDR is indeed inhomogeneous due to a central X-ray absorber that shields the X-rays along the line of sight, while they can stimulate molecular emission in the perpendicular direction. A sketch of such a situation is given in Fig.~\ref{fig:quasar}. In the presence of a torus, indeed it seems likely that X-ray emission is preferred in the direction orthogonal to the torus, and the XDR fluxes may therefore be preferentially detectable in situations where the molecular disk is observed face-on.

\subsection{APM 08279}
In the $z=3.9$ galaxy APM~08279, the usual geometric effects concerning emission from the central regions are compensated by a gravitational lens. It therefore provides an ideal test case to study the emission we might see when future telescopes like ALMA can observe the central regions with higher sensitivity and higher spatial resolution. As {expected for molecular clouds excited by X-rays}, they find that the CO line intensity increases up to the (10-9) transition. Significant flux is also detected in the HCN(5-4) line. 

Based on brightness temperature arguments, the results from high-resolution mapping and lens models from the literature, \citet{Weiss07} show that the molecular lines arise from a compact ($100-300$~pc), highly gravitationally magnified (m=60-110) region surrounding the central AGN. It is interesting to note that this amount of magnification is comparable to the increase in sensitivity due to ALMA. They can distinguish two components of the gas: A cool component with a density of $10^5$~cm$^{-3}$ and a temperature of $65$~K, on scales larger than $100$~pc, and a warm component of gas with $10^4$~cm$^{-3}$ and temperatures of $220$~K on scales smaller than $100$~pc. So far, their results do not provide indications for inhomogeneities.

We have compared the turnover for the rotational CO lines found by \citet{Weiss07} with the grid of PDR and XDR model predictions that were made publicly available by \citet{Meijerink07}. In the PDR case, we find that cloud densities of $10^5$~cm$^{-3}$ and a radiation field of $G_0\sim10^{4.75}$ with a cloud column of $\sim6\times10^{22}$~cm$^{-2}$ are required. For the XDR case, the turn-over can be explained with a cloud density of $10^{4.25}$~cm$^{-3}$, an X-ray flux of $2.8$~erg~s$^{-1}$~cm$^{-2}$ and a cloud column density of $\sim2.6\times10^{22}$~cm. In both cases, we need to require that additional cold gas is present that emits in particular in the low-lying $J$ levels to explain the observed ratio between high-$J$ and low-$J$ rotational CO lines.

The X-ray spectrum of APM~08279 has been observed with Chandra \citep{Chartas02}, indicating a luminosity of $4\times10^{46}m^{-1}$~erg~s$^{-1}$. Adopting a lens magnification of $m\sim100$, one expects a flux of $\sim354$~erg~s$^{-1}$~cm$^{-2}$ at a distance of $100$~pc for an optically thin source. However, for a density of $10^{4.25}$~cm$^{-3}$, we expect significant attenuation effects that may considerably decrease the flux in the cloud. As shown by \citet{Wada09}, shielding may locally vary by two orders of magnitude due to column density fluctuations in the torus and give rise to a peak optical depth of $\tau\sim5$ for frequencies of $3$~keV.

With ALMA, it will be possible to probe the distribution of these gas components and their kinematics in even more detail. Due to the higher sensitivity, the error bars in the flux measurement will decrease further and one may probe whether the turnover occurs at the (10-9) transition, as indicated now, or indeed even at higher-$J$ levels. In this case, one could clearly discriminate between PDR and XDR models.

\subsection{SDSS~J114816.64+525150.3}\label{quasar6}
{
The discovery of the extended source SDSS~J114816.64+525150.3 at $z=6.42$ by \citet{Walter09} has stimulated great observational interest that led to a broad collection of data on this source. Recently, \citet{Riechers09} provided a table with fluxes in the CO (3-2), (6-5) and (7-6) transitions, fluxes in the CI line and the [CII] line, as well as upper limits on five additional lines. Although the publication of such upper limits is very valuable and can provide important constraints on theoretical models, for simplicity we focus here on those lines that were actually detected. As we shall see, the interpretation of those is already complex, and a more thorough analysis would be beyond the scope of this work. The models and data empoyed here are based on the publicly available grids provided by \citet{Meijerink07}. Due to the large scales of this source, we assume that most of the flux is driven by PDR chemistry, especially because the relevant scales for XDRs are currently unresolved.}

{
To understand the complexity of these data, it is illustrative to check whether a model with one given density and a fixed parameter $G_0$ is able to explain them. For CO, the ratio between the (6-5) and the (3-2) transition is about 3.35, while the ratio between (7-6) and (3-2) is about 3.15. Although it appears that the turnover in the SED may have been reached, this is not fully clear from the data, as the uncertainty in the fluxes is $\sim10\%$. The increased intensity in the (6-5) transition requires a significant amount of soft-UV flux, while the almost flat behavior indicates that the gas should be at intermediate densities of about $\sim10^4$~cm$^{-3}$. At higher densities, the seventh rotational level would be more easily excited, which would give rise to a significant discrepancy between the (6-5) and the (7-6) transition, even for low values of $G_0$. For lower densities, on the other hand, it becomes difficult to excite these rotational levels. A reasonable fit to the CO SED can thus be obtained with $n=10^4$~cm$^{-3}$ and $G_0=10^{1.75}$.
}

{A problem arises, though, if the observed fluxes in [CII] and CI should arise from the same gas component. In this case, our model predicts a flux ratio of [CII] to CI of $\sim600$, and the intensity of [CII] and CI would be much higher than the CO intensities. We note, however, that the fine-structure lines are already optically thick at these densities, while the CO lines are optically thin. Thus, additional clouds may be present that enhance the CO intensities, while the optically thick emission in [CII] and CI is unaffected. Nevertheless, as observations show a corresponding line ratio of $17.7$, these lines should originate from a different gas component. Our models require the presence of additional gas clouds with density of $\sim10^3$~cm$^{-3}$, $G_0\sim10^{1.75}$ and large columns $N_H>10^{21}$~cm$^{-2}$, at which enough of the soft-UV flux is shielded to yield a low ratio between [CII] and CI.  This line ratio is indeed strikingly low, as PDR models generally predict larger numbers for this ratio, and also in this case the match is not perfect, but only within the $20\%$ error of the CI measurement. Although this may be an issue due to uncertain abundance ratios, a more precise determination of the flux in this line is highly desirable.}

{So far, we have postulated two gas components to explain the CO line SED and the fluxes from [CII] and CI. It is however important to cross-check whether these components may affect the line ratios that the other component should reproduce. For the CO line SED, indeed the intensity in a low-density cloud is smaller by one order of magnitude. The high-density component, however, gives rise to an intensity in [CII] that is smaller than the corresponding intensity from the other component by just a factor of $5$. To avoid that this perturbes the [CII] to CI flux ratio, we need to require that the low-density gas is more abundant than the gas at high densities. In this case, it seems however likely that the low-density gas would perturb the CO line SED significantly.}

{The most probable explanation in terms of a two-component picture is thus that indeed the low-density component is more abundant and explains the observed fine-structure lines. The impinging soft-UV flux must be moderate in order to reconcile the low ratio between these lines. To still explain the observed CO line SED, we need to require that the high-density component is at higher temperatures, due to an increased value of $G_0\sim10^3$. As one generally expects that high-density gas is exposed to a weaker radiation field due to shielding effects, this component should be spatially separated in regions of strong active star formation. Alternative scenarios are however feasible as well. For instance, the observed [CII]/CI ratio may also be produced in the presence of a weak X-ray background rather than a soft-UV field, or if a the cosmic-ray background is enhanced by a factor of 10-100 compared to the Milky Way \citep[see also][]{Meijerink06b}. }

{Of course, this model is still an oversimplification, as one may indeed expect a variety of different densities in both star-forming and more quiescent parts of the galaxy. In addition, there are theoretical uncertainties concerning the metallicity and the abundance ratios that may affect our results. Nevertheless, this model reproduces the main features in the observed fluxes and illustrates the difficulties in simultaneously modeling observations in different lines. This challenge will become more severe, as the high sensitivity of ALMA will allow us to study even more lines. On the other hand, it is certainly desirable if some of the simplest models can be discarded on such grounds. We note that such models are also predictive. For the scenario described here, we would for instance expect that also the flux in the [OI]~$63$~$\mu$m line and the [OI]~$145$~$\mu$m line originates predominantly from the low-density gas component. In this case, the ratio of these fluxes to the flux in [CII] would be $0.06-0.1$ and $8\times10^{-4}$, respectively. If the oxygen abundance is enhanced by a factor of $4$ compared to the galactic abundance, the flux in the $63$~$\mu$m line would be larger by a factor of 2.}

{
 With future ALMA measurements, we expect that the CO line SED can be probed in more detail and with higher accuracy at higher-$J$ levels, to check whether the turnover has already been reached. Apart from the observed gas component that indicates a turnover near the $J=6$ transition, a higher density component may be present that can more easily excite flux at higher-$J$ levels. The increased sensitivity and angular resolution of ALMA may help to detect spatial variations in the different fluxes, which may help to discriminate regions of intense star formation from more quiescent zones. In the center of the galaxy, ALMA can check for the presence of an X-ray dominated region. Significant progress may however be possible in the mean time, for instance by measuring additional high-$J$ CO lines or by a more accurate determination of the [CII] to CI flux ratio.}

\section{The expected number of sources}\label{deep_field}

As shown in \S~\ref{PDR}, large-scale PDR fluxes, for instance in the [CII]~$158$~$\mu$m line or the [OI]~$146$~$\mu$m line, are sufficiently large for detection in a few hours. Fluxes from the XDR originate from a smaller region, but may also give rise to a significant flux component. With an integration time of a few hours, active galaxies with a comparable brightness as NGC~1068 should be detectable in an ALMA field of view. One might therefore wonder whether a search for active galaxies, based on the before-mentioned fine-structure lines, is feasible. {Beyond that, it is of strong interest to know whether there are reasonable chances to find active galaxies with other surveys such as JWST. In such cases, ALMA could perform relevant follow-up studies that probe the central X-ray dominated regions and the gas dynamics within them. If a sufficient number of sources is obtained, ALMA can probe the feeding of black holes between redshift 6 and 10 and thus constrain models concerning the growth of supermassive black holes based on the diagnostics provided here. We therefore conclude this paper with an estimate on the high-redshift black hole population.}
%
%Of course, flux in these lines does not necessarily originate from AGNs, but is also generated in star-forming regions. Therefore, any potentially obtained source catalog needs to be analyzed using follow-up observations. This could be done with ALMA itself, by employing the largest baselines to obtain high angular resolution such that potential XDRs and PDRs can be separated and discriminated based on their line ratios \citep[see][]{Meijerink05}. An alternative might be follow-up observations by state-of-the art X-ray or radio telescopes to investigate the nature of these galaxies which could provide clear evidence for the presence of an accreting black hole.

To assess this possibility in more detail, we start with some general considerations regarding the black hole population at $z>6$. Then we discuss estimates on the number of sources and the possibility to detect them with an ALMA deep field.

\subsection{Black hole growth at high redshift}\label{growth}

\begin{figure}
\includegraphics[scale=0.4]{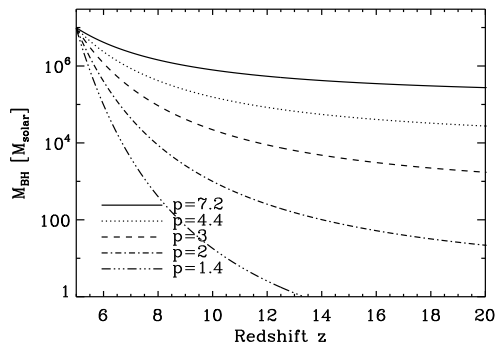}
\caption{{The average accretion history of a black hole with $10^7\ M_\odot$ at $z=5$, depending on a coefficient $p$ which is a function of the conversion of rest-mass energy to luminous energy, the Eddington ratio and the duty cycle.}}
\label{fig:massev}
\end{figure}

As shown recently by \citet{Shankar09}, a supermassive black hole with $10^9\ M_\odot$ today had, on average, $10^7\  M_\odot$ at $z=5$. Therefore, one needs to explain how supermassive black holes have accreted this mass at early times. Indeed, some of them need to accrete even faster, as $10^9\ M_\odot$ black holes have already been detected at $z>6$ \citep[e.g.][]{Fan06}. We will however focus on the more conservative case of $10^7\ M_\odot$ black holes.

We follow \citet{Shapiro05} and describe black hole growth by accretion using the formula,
\begin{equation}
\frac{dM_{\mathrm{BH}}}{dt}=\frac{M_{\mathrm{BH}}}{\tau_{\mathrm{growth}}},
\end{equation}
{where }
\begin{eqnarray}
\tau_{\mathrm{growth}}&=&0.0394\frac{(\epsilon_M/0.1)}{1-\epsilon_M}\frac{1}{\lambda_E}\frac{1}{f_{\mathrm{duty}}}~\mathrm{Gyr}\nonumber\\
&\equiv&0.0394p~\mathrm{Gyr},
\end{eqnarray}
{and where $\epsilon_M=L/\dot{M}_0c^2$ is the efficiency of conversion of rest-mass energy to luminous energy, $\lambda_E$ the Eddington ratio and $f_{\mathrm{duty}}$ the duty cycle, i. e. the fraction of quasars active at a time.  The parameter $p$ summarizes the complicated dependence on the latter parameters.  With the parameters inferred by \citet{Shankar09} for the observed high-redshift black hole population up to $z=6$, we obtain $p\sim25$.

The dependence of the black hole mass evolution in this parameter is given in Fig.~\ref{fig:massev}. For $p>7$, the black hole mass hardly evolves with redshift, implying that black holes would need to form with extremely high masses. There are, however, no theoretical models available that would explain such massive seeds, which would have comparable masses to the first galaxies at high redshift. 

It is more likely that the black hole population evolved with redshift, implying that a larger fraction of them was active and had higher Eddington ratios. For instance, a value of $p=2.5$ requires intermediate mass black holes with $10^4-10^5\ M_\odot$ to be the progenitors of the first supermassive black holes. Such scenarios have been suggested by \citet{Eisenstein95}, \citet{Koushiappas04}, \citet{Begelman06}, \citet{Spaans06}, \citet{Lodato06} and \citet{Dijkstra08}. For even smaller values of $p$, supermassive black holes could even originate from stellar progenitors. 

In this respect, it is speculated that so-called dark stars, which would be stars powered by dark matter annihilation rather than nuclear fusion \citep{Spolyar08,FreeseBodenheimer08}, could obtain even larger masses than conventional Pop.~III stars. However, the high masses suggested in these works are not confirmed by other authors \citep[see also work by ][]{Iocco08, IoccoBressan08} and may be in conflict with the observed reionization optical depth \citep{SchleicherBanerjee08a, SchleicherBanerjee08c}. 

In all of these cases, it is evident that the black hole would need to grow considerably by accretion, and that a change in the parameters describing the black hole population is required. In particular, a larger fraction of active quasars and a higher Eddington efficiency are required to obtain the black hole masses that we observe today. As pointed out by \citet{Kawakatu09}, even super-Eddington accretion may be required to obtain the observed black hole masses.

\subsection{Estimates on the number of sources}

\begin{figure}
\includegraphics[scale=0.4]{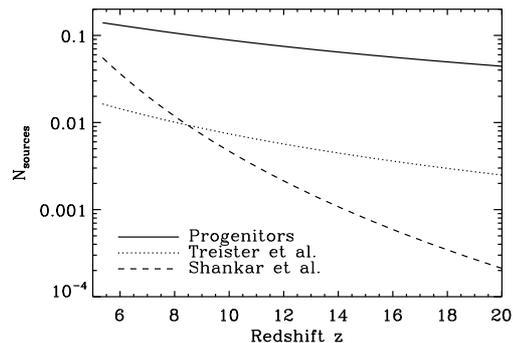}
\caption{Estimates for the number of sources in a redshift interval of $\Delta z=0.5$ within a solid angle of $(1')^2$. We show extrapolations of the results by \citet{Shankar09} and \citet{Treister09} to high redshift, for luminosities larger than $10^{44}$~erg~s$^{-1}$. In addition, we give an estimate based on the number of high-redshift black holes required in order to produce the present-day $10^9\ M_\odot$ black holes. As discussed in the text, the actual number of black holes should lie in between these estimates.}
\label{fig:further}
\end{figure}

To estimate the expected number of high-redshift black holes, we will show extrapolations of theoretical models describing the observed high-redshift black hole population, as well as estimates based on the number of $10^9\ M_\odot$ black holes in the local universe. 

To translate a given black hole number density into the number of sources in a certain field of view, we adopt a similar formalism as \citet{Choudhury07} and count the number of black holes $N(z,M_{\mathrm{BH}})$ with masses larger than $M_{\mathrm{BH}}$ in a redshift interval $[z,z+\Delta z]$ { per solid angle}. This is given as
\begin{eqnarray}
N(z,M_{\mathrm{BH}})&=&\int_z^{z+\Delta z}dz' \frac{dV}{dz' d\Omega}\int_{M_{\mathrm{BH}}}^{M_{\mathrm{max}}} d\log M'_{\mathrm{BH}} \nonumber \\
&\times& f_a \frac{dn}{d\log M_{\mathrm{BH}}},\label{counts}
\end{eqnarray}
where $dV\ dz'^{-1}\ d\Omega^{-1}$ denotes the comoving volume element per unit redshift per unit solid angle, which is given as \citep{Peebles93}
\begin{equation}
\frac{dV}{dz' d\Omega}=D_A^2 c \frac{dt}{dz}.
\end{equation}
In this expression, $D_A$ is the angular diameter distance, $c$ is the speed of light and $dt/dz=1/\left (H(z)(1+z)\right )$, where $H(z)$ is the expansion rate as a function of redshift. The term $dn/d\log M_h$ describes the number density of dark matter halos per unit mass, and the maximum black hole mass is taken as $M_{\mathrm{max}}=10^9\ M_\odot$. 

As explained above, we focus here on the possibility to search for fine-structure line emission. With a bandwith of $8$~GHz, we expect that ALMA could scan a redshift interval of $\delta z=0.2$ at $z\sim8$ in bands 6 and 7 within an integration time of a few hours. We assume here that this process is iterated until a total redshift interval of $\Delta z=0.5$ is covered, which we adopt as a fiducial redshift range in our calculation. Alternatively, one could focus on the detection of continuum flux, which would allow to scan an even larger redshift interval within a given time, but at the same time it would be more difficult to infer the redshift of source.

For our first estimate that extrapolates theoretical models of \citet{Shankar09} and \citet{Treister09}, we use their results for quasars with luminosities larger than $10^{44}$~erg~s$^{-1}$, which can be associated with a black hole mass of $10^7\ M_\odot$ for an Eddington ratio of $10\%$, as they find in their calculations. Note that \citet{Treister09} only provide numbers for Compton-thick quasars, but give a Compton-thick quasar fraction of $\sim10\%$, which we use to infer the total population. The results based on this approach are given in Fig.~\ref{fig:further} and are only approximate. 

The other approach is based on the number of $10^9\ M_\odot$ black holes in the local universe. 
As noted above, such black holes had an average mass of $10^7\ M_\odot$ at $z=5$ \citep{Shankar09}. Therefore, the comoving number density of such black holes can be estimated using the number density of today's $10^9\ M_\odot$ black holes, which is $\sim10^{-3.5}$~Mpc$^{-3}$ \citep{Shankar09}. It may be even higher, as Compton-thick black holes at $z=0$ are hard to detect \citep{Treister09}. The expected number of sources based on this estimate is shown in Fig.~\ref{fig:further}. 

One may note that the number of sources obtained in this way is larger than the value from the extrapolation of theoretical models, i. e. by \citet{Shankar09} and \citet{Treister09}. Such a behavior can be expected from our discussion in \S~\ref{growth}: As supermassive black holes should have higher duty cycles at $z>6$, the models by \citet{Shankar09} and \citet{Treister09} will naturally underestimate this population. At the same time, the estimate based on the present-day $10^9\ M_\odot$ population is also an upper limit, since not all of these black holes will be active at the same time. We therefore expect that these approaches yield an upper and a lower limit and that the actual black hole population will lie in between.

For the expected abundance of $10^7\ M_\odot$ black holes at $z=5$, we therefore adopt a fiducial number of $\sim0.08$ in a solid angle of $(1')^2$, and a number of $\sim0.03$ at $z=8$. We estimate the uncertainty to be half an order of magnitude. Based on the large fluxes found in \S~\ref{ngc1068} and the results of \citet{Spaans08}, we argue that even smaller systems with $10^6\ M_\odot$ black holes should be detectable. Calculations by \citet{Shankar09} show that the cumulative space density of high-redshift quasars increases by almost an order of magnitude if the minimum luminosity is lowered by an order of magnitude. Additional sources may be present as well, such as starburst galaxies or ULIRGS. One may therefore expect one active galaxy in a field of view of one arcmin$^2$ at $z\sim6$. 

At higher redshift, this number may however decrease significantly. Therefore, searching for new sources in an ALMA deep field seems difficult for $z>6$, and it is important that surveys as performed by JWST\footnote{http://www.jwst.nasa.gov/} provide a catalogue of high-redshift sources. {Based on the estimates discussed here, we conclude that this may be sufficient to study black hole accretion between redshifts $6$ and $10$, and perhaps even beyond.}

\section{Conclusions}\label{conclusions}
In this paper, we have explored how ALMA observations may help to extend our current knowledge concerning the observations of high-redshift quasars. An obvious result is that current observations of the starburst component can be done with higher resolution and higher sensitivity, which may therefore provide detailed local gas kinematics. Due to the higher sensitivity, it seems likely that also a number of the weaker fine-structure lines may be detected, like [NII]~$122$~$\mu$m for $z>2.5$, or [SIII]~$34$~$\mu$m and [SiII]~$35$~$\mu$m for $z>11$, providing a valuable independent probe of the gas chemistry.

Due to the combination of high resolution and high sensitivity, ALMA may however go beyond that and try to explore the centers of quasar host galaxies at high redshift. As the models presented here indicate, one may expect that the chemistry in these regions is dominated by X-ray emission from the central engine. In contrast to soft-UV photons, X-rays have high heating efficiencies and low efficiencies for dissociation. They can therefore give rise to an entirely different molecular cloud chemistry, and can excite high-$J$ CO lines well above the (10-9) transition. 

Additional probes for the central XDRs are available as well, in particular fine-structure lines like [OI]~$63$~$\mu$m, [OI]~$146$~$\mu$m and [CII]~$158$~$\mu$m. In particular for high X-ray luminosities, the fine-structure lines of [OI]~$63$~$\mu$m and [CII]~$158$~$\mu$m may provide a significant amount of flux even for gas densities of $10^4$~cm$^{-3}$. In general, the fine-structure lines become optically thick more easily than the rotational CO lines, in particular at densities of $10^5$~cm$^{-3}$. Therefore, they are almost insensitive to the column density, but quite sensitive to the X-ray flux. The CO lines, on the other hand, have a more non-trivial dependence on the X-ray flux and the column density, as in particular for high fluxes some dissociation effects will be present. If ideally both the CO lines and the fine-structure lines are observed, it may thus be possible to derive constraints both on column densities and the
local X-ray flux. The line ratios may also provide valuable information on cloud gas density and temperature, as pointed out by \citet{Meijerink07}.

We have applied our models to available data for observed XDRs. For galaxies such as NGC~1068, we expect that the CO line intensity rises continuously up to the (17-16) transition. Such high excitation is not possible in PDR models \citep{Spaans08}. A detection of such lines therefore provides a unique diagnostic for an XDR. 

We consider another example, the lensed galaxy APM~08279. There, the situation is not fully clear and is currently consistent with the presence of an XDR, which should also be expected due to the high X-ray luminosity, but also with the presence of a PDR with a radiation field of $G_0=10^{4.5}$.  To some degree, this is due to uncertainties in the measured line fluxes. We expect that improved measurements with ALMA may help to distinguish these scenarios further and probe the local gas motions in more detail.

We finally discussed the chances to find additional sources with ALMA in a survey that scans a few arcmin$^2$. General considerations regarding the average accretion history indicate that quasar duty cycles should be larger at high redshift, so that black holes can grow to the required masses at $z=6$. We therefore expect more sources compared to what one expects from a naive extrapolation of the observed quasar population. Our estimates indicate that about one source per arcmin$^2$ may be present near $z\sim6$. At larger redshift, the number of sources may decrease significantly and one may need to rely on follow-up observations of sources  detected by JWST. Due to the potentially high spatial resolution, the transport of gas to the center of the host galaxy may be probed for the first time in such galaxies even at $z\sim10$.

{We finally note that the diagnostics considered in this paper, in particular the high-$J$ CO lines, may not only be important for high-redshift observations with ALMA, but also for future space-borne observations in the local universe with SPICA or FIRI. }

\begin{acknowledgements}
We thank Robi Banerjee, {Max Camenzind}, Wilfred Frieswijk, Simon Glover, Edo Loenen, Rowin Meijerink, Padelis Papadopoulos, Dieter Poelman and Fabian Walter for valuable discussions on the topic. The research leading to these results has received funding from the European Community's Seventh Framework Programme (/FP7/2007-2013/) under grant agreement No 
229517. {RSK thanks the German Science Foundation (DFG) for support via the Emmy Noether grant KL 1358/1.} DRGS and RSK also acknowledge subsidies  from the DFG SFB 439 {\em Galaxies in the Early Universe}, and grant KL 1358/10 {under the Priority Programme 1177 {\em"Witnesses of Cosmic History:  Formation and evolution of black holes, galaxies and their environment"} of the German Science Foundation}, as well as via the FRONTIER program of Heidelberg University. In addition, R.S.K.\   
thanks for subsidies from the German {\em Bundesministerium f\"{u}r  
Bildung und Forschung} via the ASTRONET project STAR FORMAT (grant  
05A09VHA) and from the {\em Landesstiftung Baden-W{\"u}rttemberg} via  
their program International Collaboration II. {We thank the anonymous referee for valuable comments that helped to improve the manuscript.}

\end{acknowledgements}

%\bibliographystyle{apj} %z.b. ApJ Style
%\bibliography{astro}

\end{document}